\documentclass[11pt]{iopart}
\usepackage{bm,bbm}	
\usepackage{graphicx} 
\usepackage[citecolor=black,linkcolor=black]{hyperref} 
\usepackage{array} 
\usepackage{color}

\begin{document}

\title{Nonequilibrium quantum heat transport between structured environments}
\author{Graeme Pleasance$^1$$^2$ and Francesco Petruccione$^1$$^2$$^3$}
\address{$^1$ Department of Physics, University of Stellenbosch, Stellenbosch, 7600, South Africa}
\address{$^2$ National Institute for Theoretical and Computational Sciences (NITheCS), South Africa}
\address{$^3$ School of Data Science and Computational Thinking, University of Stellenbosch, Stellenbosch, 7600, South Africa}
\ead{gpleasance1@gmail.com}

\begin{abstract}

We apply the hierarchical equations of motion technique to analyzing nonequilibrium heat transport in a spin-boson type model, whereby heat transfer through a central spin is mediated by an intermediate pair of coupled harmonic oscillators. The coupling between each pair of oscillators is shown to introduce a localized gap into the effective spectral densities characterizing the system--oscillator--reservoir interactions. Compared to the case of a single mediating oscillator, we find the heat current to be drastically modified at weak system-bath coupling. In particular, a second-order treatment fails to capture the correct steady-state behavior in this regime, which stems from the $\lambda^4$-scaling of the energy transfer rate to lowest order in the coupling strength $\lambda$. This leads naturally to a strong suppression in the steady-state current in the asymptotically weak coupling limit. On the other hand, the current noise follows the same scaling as in the single oscillator case in accordance with the fluctuation-dissipation theorem. Additionally, we find the heat current to be consistent with Fourier's law even at large temperature bias. Our analysis highlights a novel mechanism for controlling heat transport in nanoscale systems based on tailoring the spectral properties of thermal environments. 

\end{abstract}

\section{Introduction}

Understanding and controlling energy transport mechanisms at the microscopic level is of critical importance to the development of nanoscale quantum technologies. Nanoscale devices that exploit the steady-state transfer of energy or particles (electrons, photons, phonons) through a quantum system---including thermal rectifiers \cite{Segal2005,Segal2005a,Ruokola2009}, heat engines \cite{Rossnagel2016,Newman2017,Klatzow2019,Wiedmann2020,Latune2023,Bhattacharjee2021} and refrigerators \cite{Levy2012,Correa2014,Maslennikov2019,Ivander2022}---have become an increasingly active topic of research due to their potential technological applications \cite{Dubi2011}. From a more fundamental perspective, the study of such systems also plays an important role in advancing our understanding of quantum thermodynamics \cite{Vinjanampathy2016,Goold2016}. Despite most activity in these fields remaining largely theoretical, there has been significant progress in recent years towards the realization of quantum thermal machines and, in particular, circuit quantum thermodynamics \cite{Pekola2015,Ronzani2018}, with the thermal diode effect having recently been demonstrated in superconducting quantum circuits \cite{Senior2020}. The growing relevance of these technologies thus calls for deeper insights into ways of controlling energy transport at the nanoscale.

For boundary-driven systems, i.e. those in which a heat current is driven by local interactions with thermal baths \cite{Landi2022}, the nonequilibrium spin-boson model (NESB) provides one of the simplest yet most widely applied models of energy transport exhibiting a rich phenomenology. The model consists of a two-level system (spin-$\frac{1}{2}$) coupled to two Ohmic baths at different temperatures, and has found widespread applications to anharmonic molecular junctions \cite{Segal2006,Segal2006a}, cold atoms, and biological systems \cite{Gilmore2005}. To date, numerous analytical and numerical techniques have been employed to study the NESB, each providing varying levels of insight into its heat transport characteristics. These include perturbative master equations \cite{Segal2005,Segal2006}, nonequilibrium Green's function methods (NEGF) \cite{Wang2013,Agarwalla2017}, path integral techniques \cite{Simine2013,Carrega2016,Aurell2018,Aurell2020}, the reaction-coordinate method \cite{AntoSztrikacs2021,AntoSztrikacs2022}, the hierarchical equations of motion (HEOM) \cite{Kato2015,Kato2016,Cerrillo2016,Song2017,Kato2018,Duan2020}, and the time-evolving matrix product operator (TEMPO) technique \cite{Chen2023}, among others. Perturbative master equations constitute a standard tool of open quantum systems theory \cite{Breuer2002} and have been shown to capture resonant heat transfer in the NESB up to second-order in the system-bath coupling. Alternatively, the (nonequilibrium) non-interacting blip approximation (NIBA) \cite{Leggett1987,Nicolin2011,Segal2014,Weiss2011}, nonequilibrium polaron transformed Redfield equation (PT-RE) \cite{Wang2015,Wang2017,Liu2018}, and HEOM \cite{Cerrillo2016,Duan2020} have been used to treat the full counting statistics \cite{Esposito2009} of the NESB beyond second-order, demonstrating turnover behavior in the steady-state currents under strong system-bath interactions.

Recently, an extension to the NESB was proposed and studied in \cite{AntoSztrikacs2022,Duan2020,Novais2005,Palm2018}, where a unitary rotation is applied to one of the interaction terms so as to make the system-bath coupling operators noncommutative. This feature introduces a source of asymmetry into the model that facilitates two competing transport mechanisms: (i) one in which heat transfer occurs between baths through direct excitation of the system, termed system currents, and (ii) another pathway independent of the system in which a heat flow occurs directly between the baths. Notably, since the inter-bath current scales \textit{quartically} with respect to the system-bath coupling parameter $\lambda$ as $\lambda\rightarrow0$, it was established that a second-order approach is insufficient to capture the correct steady-state behavior in cases where the system current is zero \cite{AntoSztrikacs2022,Duan2020}. At the same time, other extensions to the NESB were examined in \cite{Aurell2021,Yamamoto2021,Xu2021} to study heat transport through superconducting circuits, consisting of a transmon qubit and pair of resonators in a resonator-qubit-resonator configuration (quantum Rabi model) embedded between two thermal reservoirs \cite{Reuther2010,Ronzani2018}. A similar setup whereby the reservoirs are `filtered' through an intermediate oscillator has also been considered in the context of autonomous thermal machines \cite{GelbwaserKlimovsky2014}.

\begin{figure}[t!]
\centering
\includegraphics[width=\textwidth]{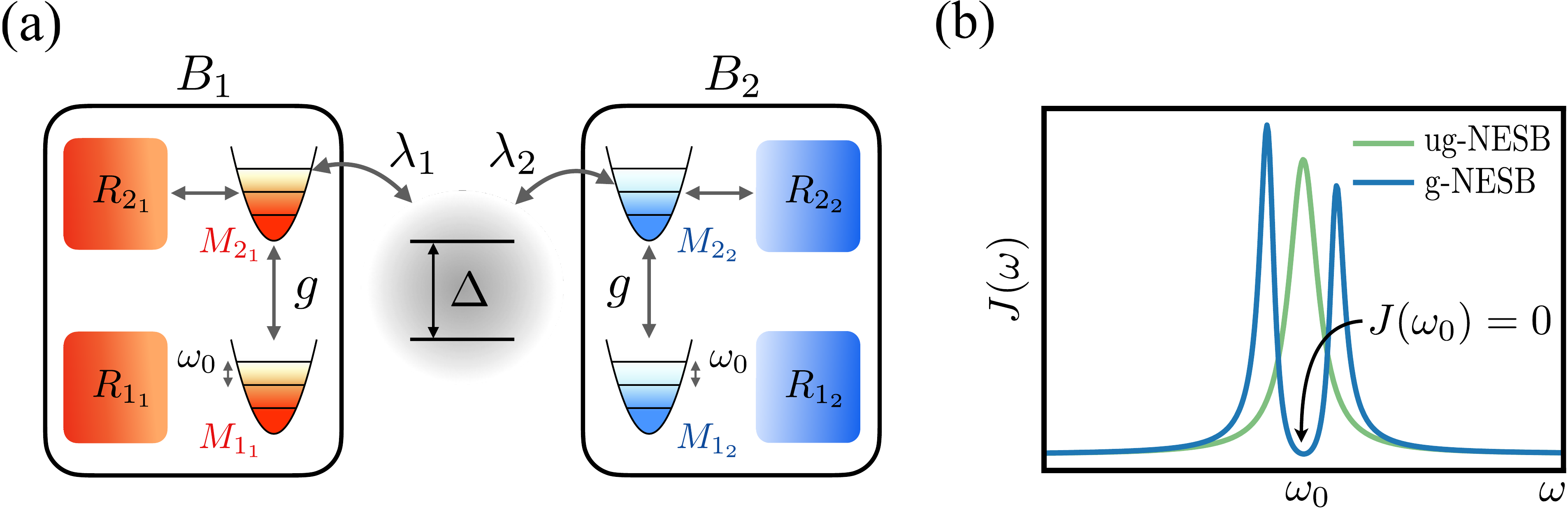}
\caption{\label{fig:1} (a) Schematic of nonequilibrium heat transport model with structured baths. The system $S$ couples to hot and cold thermal reservoirs $R$ with Ohmic spectral densities (\ref{eq:J_ohmic}) through an intermediate pair of coupled oscillators $M$. Each oscillator-pair and their reservoirs are assumed to be intialized in a composite Gibbs state $\rho_B(0) = \bigotimes_{j=1,2}e^{-\beta_jH_{B_j}}/Z_{B_j}$. (b) Plots of the bath spectral densities corresponding to the ungapped NESB (ug-NESB) and gapped NESB.}
\end{figure}

Motivated by these studies, in this paper we investigate the transport properties of a different extended NESB---the \textit{gapped nonequilibrium spin-boson model} (g-NESB)---where the usual Ohmic baths are augmented by a pair of coupled harmonic oscillators (see figure \ref{fig:1}). To our knowledge, this model has not previously been proposed or studied in the context of heat transport. We analyze the model in both the transient and steady-state regimes using the numerically exact HEOM \cite{Tanimura1989,Tanimura1990,Ishizaki2005,Tanimura2006,Ishizaki2009,Tanimura2020}. This enables us to capture higher-than-second-order effects at weak system-bath coupling, based on exact expressions for the average heat current and noise obtained via a generating functional approach \cite{Kato2016,Chen2023}. Furthermore, we directly compare its transport properties to the quantum Rabi model of \cite{Aurell2021,Yamamoto2021}, which is recovered as a special case of the g-NESB when the inter-mode coupling between oscillators is set to zero.

Our key findings are summerized by the following: (i) in the nonequilibrium steady-state the heat current exhibits a $\lambda^4$-scaling to lowest order in $\lambda$, contrasting with the $\lambda^2$-scaling of the quantum Rabi model. This reveals that the addition of a second oscillator considerably suppresses the stationary current in the weak coupling limit. (ii) The steady-state noise in both models exhibits the same $\lambda^2$-scaling to lowest order in $\lambda$ and---unlike the currents---does not exhibit turnover behavior at strong system-bath coupling. (iii) The steady-state current of the g-NESB obeys Fourier's law at large temperature bias outside the linear response regime. 

Additionally, to elucidate the quartic scaling behavior we construct perturbative expressions for the stationary current and noise based on the HEOM. We find that since the spectral densities of the effective baths exhibit a localized gap at the oscillator frequencies, the leading-order contribution to the system current vanishes when the system tunnelling frequency is resonant with the gap. Hence, in this case a second-order treatment fails to describe the correct physics even in the asymptotically weak coupling limit. However, it should be emphasized that the quartic scaling here stems from a different transport mechanism to that explored in \cite{AntoSztrikacs2022,Duan2020}, which as mentioned above, is due to the noncommutativity of the system-bath coupling operators. 

The remainder of the paper is organized as follows. In section \ref{sec:2}, we introduce the models under study and derive an analytic expression for the spectral densities of the effective baths. Section \ref{sec:3} outlines technical details relating to the implementation of the HEOM, as well as exact and perturbative expressions for the relevant quantities 
derived within this framework. In section \ref{sec:4}, we present numerical simulations of both the heat current and its noise in the transient and steady state regimes, comparing our results to those obtained from the corresponding quantum Rabi model \cite{Aurell2021,Yamamoto2021}. Finally, a summary and conclusions is provided in section \ref{sec:5}. 

\section{Model}\label{sec:2}

We consider the model of an unbiased two-level system coupled to two structured bosonic baths $B_j$, which are initially in equilibrium at temperatures $T_j$, depicted schematically in figure \ref{fig:1}. Each bath consists of a pair of coupled harmonic oscillators $M_{1_j}$ and $M_{2_j}$ that linearly interact with both the system and an additional thermal reservoir. The Hamiltonian of the system reads 
\begin{equation}\label{eq:H_S}
	H_S = \frac{\Delta}{2}\sigma_x,
\end{equation}
with $\Delta$ a tunnelling frequency and $\sigma_{x,y,z}$ the Pauli spin-$\frac{1}{2}$ operators. Throughout we choose natural units such that $\hbar=k_B=1$. The Hamiltonians describing each pair of harmonic oscillators and system-oscillator coupling are given by ($i=1,2$) 
\begin{eqnarray}
	H_{M_j} &= \omega_0\sum_ia^{\dagger}_{i_j}a_{i_j} + gX_{1_j}X_{2_j}, \\
	H_{I_j} &= \lambda_j\sigma_z\otimes X_{2_j}.
\end{eqnarray}
Here, $a_{i_j}$ ($a^{\dagger}_{i_j}$) is the annihilation (creation) operator of the $i$th oscillator in the $j$-th bath with frequency $\omega_0$, and $X_{i_j} = a_{i_j}+a^{\dagger}_{i_j}$. The parameters $\lambda_j$ and $g$ denote the coupling strength of $S$ to the second oscillator of the $j$-th bath and the coupling between oscillator pairs, respectively. We emphasize that the system is only directly coupled to \textit{one} of the oscillators in each pair and we assume the inter-mode coupling satisfies $g\ll\omega_0$. 

In addition to the coupling between oscillator pairs, each oscillator interacts with its own thermal reservoir $R_{i_j}$. As such, the Hamiltonian of each bath $B_j$ can be written as
\begin{equation}\label{eq:H_B} 
	H_{B_j} = H_{M_j} + H_{R_j} + H_{MR_j} + H_{C_j},
\end{equation}
where 
\begin{eqnarray}
	H_{R_j} &= \sum_{i,k}\tilde{\omega}_{k_j}\tilde{b}^{\dagger}_{i,k_j}\tilde{b}_{i,k_j}, \\
	H_{MR_j} &=\sum_iX_{i_j}\otimes\sum_{k}\tilde{g}_{i,k_j}(\tilde{b}_{i,k_j}+\tilde{b}^{\dagger}_{i,k_j}),
\end{eqnarray}
and $\tilde{b}_{i,k_j}$ ($\tilde{b}^{\dagger}_{i,k_j}$) are the corresponding annihilation (creation) operators of the reservoir $R_{i_j}$. The spectral density of each reservoir is defined as $\tilde{J}_{i_j}(\omega)=\pi\sum_{k}\tilde{g}^2_{i,k_j}\delta(\omega-\tilde{\omega}_{k_j})$. By convention, we have also added a renormalization term $H_{C_j}$ to each bath Hamiltonian (\ref{eq:H_B}), which accounts for the shift in the minimum of each oscillator potential caused by the oscillator-reservoir coupling \cite{Weiss2011,Caldeira1983}. These terms are explicitly given by
\begin{equation}\label{eq:renorm}
	H_{C_j} = \sum_iX^2_{i_j}\sum_{k_j}\frac{\tilde{g}^2_{i,k_j}}{\tilde{\omega}_{k_j}} = \sum_iX^2_{i_j}\int^{\infty}_0\frac{d\omega}{\pi}\frac{\tilde{J}_{i_j}(\omega)}{\omega}, 
\end{equation}
so that the full Hamiltonian of the model reads
\begin{equation}\label{eq:H}
	H_{\rm tot} = H_S + \sum_j(H_{B_j} + H_{I_j}). 
\end{equation}
For later discussion we refer to the $g=0$ case as the \textit{ungapped nonequilibrium spin-boson model} (ug-NESB).

\subsection{Equivalent nonequilibrium spin-boson model}

Our goal is to understand how the coupling between each pair of intermediary oscillators influences the properties of the heat flux through the system. To achieve this, we focus on a path integral treatment in which dissipative effects from each of the baths on the system are prescribed through the Feynman-Vernon influence functional $\mathcal{F}_j[q,q']$ \cite{Breuer2002,Weiss2011,Feynman1963}. The influence functional appears in the propagator $\mathcal{J}(q_f,t|q_i,0)$ describing the time evolution of the reduced system under the effect of the baths, 
\begin{equation}\label{eq:propagator}
	\mathcal{J}(q_f,t|q_i,0) = \int\mathcal{D}q\int\mathcal{D}q'\,e^{i(S_S[q] - S_S[q'])}\prod_j\mathcal{F}_j[q,q'],
\end{equation}
where $S_S[q]$ is the classical system action, and $q$, $q'$ label the forward and backward system paths. Specifically, these paths represent the possible spin trajectories connecting the boundary states $q(t)=q_f$ and $q(0)=q_i$, with $q\in [-1,1]$ labelling the possible spin values in the eigenbasis of the system Hamiltonian (\ref{eq:H_S}). 

The procedure to derive a closed form expression for $\mathcal{F}_j[q,q']$ relies on the functional integrals (\ref{eq:propagator}) being Gaussian. For non-interacting harmonic environments in thermal states, this is indeed the case and the influence functional reduces to the well-known exponential form
\begin{equation}
\mathcal{F}_j[q,q']=e^{i\Phi_j[q,q']},
\end{equation}
with $\Phi_j[q,q']$ represented as a double time integral over the forward and backward system paths [see (\ref{eq:FV})]. Since for non-interacting environments the influence functionals can be determined exactly, it will prove convenient for us to then employ a normal mode representation of the baths $B$ in order to facilitate HEOM calculations of the heat currents\footnote{We emphasize that such a representation can be freely chosen for the baths $B$ provided that (i) the system-bath interaction remains linear in annihilation and creation operators, (ii) the bath Hamiltonian remains quadratic in these operators, and (iii) the initial bath states are thermal states at temperatures $T_j$.}. The Hamiltonians of the system and baths in such a representation are written as  
\begin{eqnarray}
	H_{B_j}&=\sum_k\omega_{k_j}b^{\dagger}_{k_j}b_{k_j}, \label{eq:H_B_eff} \\
	H_{I_j}&= \sigma_z\otimes\sum_kg_{k_j}(b_{k_j}+b^{\dagger}_{k_j}) \equiv \sigma_z\otimes B_j. \label{eq:H_I_eff}
\end{eqnarray}
Here, $b_{k_j}$ ($b^{\dagger}_{k_j}$) is the bosonic annihilation (creation) operator for the $k$-th normal mode of the $j$-th bath with frequency $\omega_{k_j}$. $g_{k_j}$ denote the couplings between each normal mode oscillator of the baths and the system. Given that the total initial state is factorized as
\begin{equation}\label{eq:therm_state}
	\rho_{\rm tot}(0) = \rho_S(0)\bigotimes_j\frac{e^{-\beta_jH_{B_j}}}{Z_{B_j}},
\end{equation}
with $T_j=1/\beta_j$ the bath temperature and $Z_{B_j}={\rm Tr}[e^{-\beta_jH_{B_j}}]$, the corresponding path integral expressions for the influence functionals $\mathcal{F}_j[q,q']$ are \cite{Feynman1963}
\begin{equation}\label{eq:FV}
	\fl \qquad \mathcal{F}_j[q,q'] = \exp\left(-\int^t_0d\tau\big[q(\tau) - q'(\tau)\big]\int^{\tau}_0ds\big[C_j(\tau-s)q(s) - C^*_j(\tau-s)q'(s)\big]\right), 
\end{equation}
where
\begin{equation}\label{eq:BCF}
	C_j(t) = \frac{1}{\pi}\int^{\infty}_0d\omega\,J_j(\omega)\Big[\coth\left(\frac{\beta_j\omega}{2}\right)\cos(\omega t) - i\sin(\omega t)\Big],
\end{equation}
are the bath correlation functions, and 
\begin{equation}\label{eq:J_eff}
	J_j(\omega)=\pi\sum_{k}g^2_{k_j}\delta(\omega-\omega_{k_j})
\end{equation}	
are the spectral densities characterizing the interaction between the effective baths $B$ and system $S$. 

Our task now is to derive an analytic expression for the spectral densities $J_j(\omega)$ through a suitable choice of $\tilde{J}_{i_j}(\omega)$. To this end, we rely on a similar ansatz employed by Garg \etal \cite{Garg1985}, in which a single harmonic oscillator interacting with an Ohmic reservoir was shown to map to an effective harmonic bath with Lorentzian spectral density \cite{IlesSmith2014,IlesSmith2016,Strasberg2016}. We therefore choose each of the spectral densities $\tilde{J}_{i_j}(\omega)$ characterizing the $M-R$ interactions to be Ohmic, 
\begin{equation}\label{eq:J_ohmic}
	\tilde{J}_{i_j}(\omega) = \frac{2\gamma_{i_j}\omega\omega^2_c}{\omega^2 + \omega^2_c},
\end{equation}
where $\gamma_{i_j}$ is a dimensionless coupling constant, and $\omega_c$ a cutoff frequency. Notice that our choice of cutoff function is different from the usual exponential-type used in references \cite{Garg1985,IlesSmith2014}: this is motivated by fact that it not only gives an effective spectral density which is practical to implement with the HEOM for $\omega_c\rightarrow\infty$, but also removes the dependence of $J_j(\omega)$ on the renormalization terms in (\ref{eq:H_B}) \cite{Correa2019}. The exact details underpinning the transformation between representations and the derivation of the corresponding spectral density (\ref{eq:J_eff}) are included in \ref{appen:A}.

Assuming the first oscillator of each bath is decoupled from its respective reservoir, $\gamma_{1_j}=0$, the effective spectral density is given by the following analytic function of $\omega$,
\begin{equation}\label{eq:J_eff_result}
	J_j(\omega) = \frac{2\lambda^2_j(\omega^2-\omega^2_0)^2\omega_0\Gamma_j\omega}{[(\omega^2-\omega^2_0)^2-4\omega^2_0g^2]^2 + (\omega^2-\omega^2_0)^2\Gamma_j^2\omega^2}.
\end{equation}
This expression fully characterizes the interaction between the system and baths in terms of the oscillator parameters $\omega_0$, $g$, $\lambda_j$ and $\Gamma_j = 4\gamma_{2_j}\omega_0$; in particular, if $g$ is set to zero so that interactions with the reservoirs are mediated through a single oscillator, we recover a Lorentzian spectral density in accordance with \cite{Garg1985}. Our result therefore extends the ansatz of Garg \etal \cite{Garg1985} to a more general, non-Lorentzian environment, which admits a representation in terms of pairs of damped oscillators.

A key feature of (\ref{eq:J_eff_result}) compared to the single oscillator case is the localized gap introduced into the bath spectrum at the oscillator frequency $\omega_0$ \cite{Kofman1994}. This is depicted in figure \ref{fig:1}(b), where $J_j(\omega)$ is plotted alongside the corresponding Lorentzian spectral density obtained for $g=0$. We investigate below how this feature plays a direct role in modulating the heat flux when the system tunnelling energy is tuned to the gap.

The bath correlation functions associated with $J_j(\omega)$ may be readily obtained by way of contour integration. The resulting expressions are given by 
\begin{equation}\label{eq:corr_func}
	C_j(t) =\sum^{\infty}_{k=0}c_{jk}e^{-iz_{jk}t} , \qquad t\geq0,
\end{equation}
where $z_{jk}$ are complex exponents derived from the pole locations of the bath spectra in the lower half complex plane, and $c_{jk}$ are determined from their corresponding residues. To implement the HEOM in practice it is necessary to truncate the expansion (\ref{eq:corr_func}) to a finite number of terms. To this end, we proceed to decompose the bath correlations as 
\begin{equation}
	C_j(t)=C^0_j(t)+M_j(t), 
\end{equation}
where $C^0_j(t)$ comprises a finite sum over the pole contributions from $J_j(\omega)$, and $M_j(t)$ an infinite sum over the Matsubara frequencies $\nu^j_n = \frac{2\pi n}{\beta_j}$ ($n\in\mathbbm{Z}^+$):   
\begin{equation}\label{eq:corr_Mats}
	M_j(t) = -i\frac{2}{\beta_j}\sum^{\infty}_{n=1}J_j(-i\nu^j_{n})e^{-\nu^j_nt}, \qquad t\geq0.
\end{equation}
Since terms with exponents $\nu^j_{n}t\gg1$ make almost no contribution to $C_j(t)$ over the relevant time scales, we may neglect them so that $C_j(t) \approx \sum^{K_j}_{k=0}c_{jk}e^{-iz_{jk}t}$. Given also that the Matsubara frequencies $\nu^j_{n}$ are proportional to $T_j$, one can expect the number of terms needed to accurately approximate (\ref{eq:corr_func}) to grow with decreasing bath temperature (in fact, in the limit $\beta_j\rightarrow\infty$ the number explodes since the sum in (\ref{eq:corr_Mats}) converges to an integral). In all numerical simulations we fix the bath temperatures to be large enough so that truncating $M_j(t)$ to a single term produces converged results. 

\section{Methods}\label{sec:3}
\subsection{Hierarchical equations of motion}

Based on the finite expansion of the correlation functions (\ref{eq:corr_func}), the system-bath interactions may be treated nonperturbatively using the HEOM technique  \cite{Tanimura1989,Tanimura1990,Ishizaki2005,Tanimura2006,Ishizaki2009,Tanimura2020}. In this approach, the non-Markovian effects of the baths $B_j$ on $S$ are encoded in a collection of auxiliary density operators (ADOs). These together with the reduced system density matrix $\rho_S(t)={\rm Tr}_B[\rho_{\rm tot}(t)]$ obey a set of coupled equations of motion. The ADOs in the path integral representation are defined as 
\begin{equation}\label{eq:ADOs}
	\fl \rho_{\vec{n},\vec{m}}(t) = \int\mathcal{D}q\int\mathcal{D}q'\,e^{i(S_S[q] - S_S[q'])}\Bigg(\Bigg\{\prod_j\prod_{k}\Theta_{jk}[q]^{n_{jk}}\Theta^*_{jk}[q']^{m_{jk}}\Bigg\}\prod_j\mathcal{F}_j[q,q']\Bigg)\rho_{S,i}(0),
\end{equation}
where
\begin{equation}
	\Theta_{jk}[q] = -ic_{jk}\int^t_0d\tau\,e^{-iz_{jk}(t-\tau)}q(\tau).
\end{equation}
Here, each of the ADOs is labelled by a pair of multi-index vectors $\{\vec{n},\vec{m}\} = \{(n_{11},...)^T, (m_{11},...)^T\}$, with $(n_{jk}, m_{jk})\in\{0,...,N^{\rm max}_j\}$, and $N^{\rm max}_j$ being a maximum cut-off that truncates the depth of the hierarchy associated with each bath. The reduced system density matrix is obtained as the first member of each hierarchy with the index values $|\vec{n}|=|\vec{m}|=0$, i.e. $\rho_S(t) = \rho_{\vec{0},\vec{0}}$, and $|\vec{n}|=\sum_{jk}n_{jk}$. The ADOs corresponding to the $L$-level of the hierarchy are those with indices satisfying $|\vec{n}|+|\vec{m}|=L$. Furthermore, the length of each of the vectors $\vec{n}$ and $\vec{m}$ is determined as the sum $K_1+K_2$ of the number of terms kept in the expansion of each bath correlation function. 

If we further introduce a set of scaled ADOs 
\begin{equation}\label{eq:scaled_ADOs}
\hat{\rho}'_{\vec{n},\vec{m}} = \hat{\rho}_{\vec{n},\vec{m}}/\bigg(\prod_j\prod_k\sqrt{c^{n_{jk}}_{jk}c^{*m_{jk}}_{jk}n_{jk}!m_{jk}!}\,\bigg)
\end{equation}
in accordance with \cite{Shi2009}, then the HEOM can be derived in the form (the hat indicates a Liouville superoperator representation)
\begin{equation}\label{eq:HEOM}
	\fl \quad \frac{\partial\hat{\rho}'_{\vec{n},\vec{m}} }{\partial t}= -i\bigg(\mathcal{L}_S + \sum_{j=1,2}\sum^{K_j}_{k=0}\big(z_{jk}n_{jk} - z^*_{jk}m_{jk}\big)\bigg)\hat{\rho}'_{\vec{n},\vec{m}} - i\sum_{j=1,2}\sum^{K_j}_{k=0}\big(\mathcal{L}^+_{jk} + \mathcal{L}^-_{jk}\big)\hat{\rho}'_{\vec{n},\vec{m}}. 
\end{equation}
Here, $\mathcal{L}_S\hat{\rho}_{\vec{n},\vec{m}}=[H_S,\hat{\rho}_{\vec{n},\vec{m}}]$ is the system Liouvillian, and
\numparts
\begin{eqnarray}
	\mathcal{L}^+_{jk}\hat{\rho}'_{\vec{n},\vec{m}} &= \sqrt{c_{jk}(n_{jk}+1)}\big[\sigma_z,\hat{\rho}'_{\vec{n}^+_{jk},\vec{m}}\big] + \sqrt{c^*_{jk}(m_{jk}+1)}\big[\sigma_z,\hat{\rho}'_{\vec{n},\vec{m}^+_{jk}}\big], \\
	\mathcal{L}^-_{jk}\hat{\rho}'_{\vec{n},\vec{m}} &= \sqrt{c_{jk}n_{jk}}\,\sigma_z\hat{\rho}'_{\vec{n}^-_{jk},\vec{m}} - \sqrt{c^*_{jk}m_{jk}}\,\hat{\rho}'_{\vec{n},\vec{m}^-_{jk}}\sigma_z,
\end{eqnarray}
\endnumparts
are superoperators that relate ADOs with index $(\vec{n},\vec{m})$ to those higher ($+$) or lower ($-$) in the hierarchy with indices $n_{jk}$ and $m_{jk}$ incremented by one, i.e. $\vec{n}^{\pm}_{jk}=(n_{11},...,n_{jk}\pm1,...)^T$.

\subsection{Heat currents}

We now discuss the calculation of the relevant thermodynamic quantities within the HEOM framework. Introducing the current operators $I_{B_j}=i[H_{B_j},H_{I_j}]$ and $I_{S_j}=-i[H_S,H_{I_j}]$, we identify heat with changes in the average energy of each bath \cite{Esposito2010,Kato2016}, where
\begin{equation}\label{eq:bhc}
	 \langle I_{B_j}\rangle = -\frac{d}{dt}\langle H_{B_j}(t)\rangle = \frac{d}{dt}\langle H_{I_j}(t)\rangle + \langle I_{S_j}\rangle,
\end{equation}
defines the mean heat current for the $j$-th bath, and 
\begin{equation}\label{eq:shc}
	\langle I_{S_j}\rangle = -i\langle [H_S(t),H_{I_j}(t)]\rangle,
\end{equation}
is the corresponding system heat current. Note that the time dependence of operators inside the averages $\langle...\rangle$ is generated in the Heisenberg picture. The minus sign in (\ref{eq:bhc}) indicates that a positive heat current is associated with a net flow of energy from the bath into the system. Using a generating functional approach \cite{Kato2016,Chen2023,Gribben2022}, the heat currents may be evaluated directly in terms of the ADOs extracted from the first level of the HEOM \cite{Song2017,Zhu2012}, as detailed in \ref{appen:B}: 
\begin{eqnarray}
	\langle I_{S_j}\rangle &= -\Delta\sum_k{\rm Tr}[\sigma_y(\hat{\rho}_{\vec{0}^+_{jk},\vec{0}} + \hat{\rho}_{\vec{0},\vec{0}^+_{jk}})], \label{eq:shc_heom} \\
	\langle I_{B_j}\rangle &=  -i\sum_kz_{jk}{\rm Tr}[\sigma_z\hat{\rho}_{\vec{0}^+_{jk},\vec{0}}] + i\sum_kz^*_{jk}{\rm Tr}[\sigma_z\hat{\rho}_{\vec{0},\vec{0}^+_{jk}}]. \label{eq:bhc_heom}
\end{eqnarray}
Note that in the nonequilibrium steady state $\rho_{\rm tot}\rightarrow \rho^{ss}_{\rm tot}$ we have $\partial_t\langle H_{I_j}(t)\rangle=0$, and so $\langle I_{S_j}\rangle_{ss} = \langle I_{B_j}\rangle_{ss}$. From the first law of thermodynamics, $\langle I_{B_1}\rangle_{ss}=-\langle I_{B_2}\rangle_{ss}$, it follows that the symmetrized heat current obtained in the asymptotic limit $t\rightarrow\infty$,
\begin{equation}\label{eq:I_ss}
	I_{ss} \equiv \frac{1}{2}(\langle I_{B_1}\rangle_{ss} - \langle I_{B_2}\rangle_{ss}),
\end{equation}
is positive by definition. According to the second law of thermodynamics, the bath currents must also satisfy
\begin{equation}
	-\sum_j\beta_j\langle I_{B_j}\rangle_{ss}\geq 0,
\end{equation}
such that the rate of entropy production in the steady state is non-negative. 

Higher order moments of the operators $I_{S_j}$ and $I_{B_j}$, which represent fluctuations in the heat currents, may also be calculated via the same generating functional approach. In particular, based on the general expressions for the $n$-th order moments derived in \ref{appen:B}, the noises relating to the currents $\langle I_{S_j}\rangle$ and $\langle I_{B_j}\rangle$ are determined as \cite{Song2017}
\begin{eqnarray}
	\fl \qquad \quad  \langle I^2_{S_j}\rangle &= \Delta^2\Big(C_j(0) + \sum_{k,k'}{\rm Tr}[\hat{\rho}_{\vec{0}^{+}_{jk,jk'},\vec{0}} + 2\hat{\rho}_{\vec{0}^+_{jk},\vec{0}^+_{jk'}} + \hat{\rho}_{\vec{0},\vec{0}^{+}_{jk,jk'}}]\Big), \label{eq:shc_so} \\
	\fl \qquad \quad \langle I^2_{B_j}\rangle &= C'_j(0) - \sum_{k,k'}{\rm Tr}[z_{jk}z_{jk'}\hat{\rho}_{\vec{0}^{+}_{jk,jk'},\vec{0}} - 2z_{jk}z^*_{jk'}\hat{\rho}_{\vec{0}^+_{jk},\vec{0}^+_{jk'}} + z^*_{jk}z^*_{jk'}\hat{\rho}_{\vec{0},\vec{0}^{+}_{jk,jk'}}], \label{eq:bhc_so}
\end{eqnarray}
where $\vec{0}^+_{jk,jk'}$ denotes ADOs with indices $0_{jk}$ and $0_{jk'}$ raised by one, and 
\begin{equation}
C'_j(t) = \frac{1}{\pi}\int^{\infty}_0d\omega\,\omega^2J_j(\omega)\Big[\coth\left(\frac{\beta_j\omega}{2}\right)\cos(\omega t) - i\sin(\omega t)\Big].
\end{equation}
Hence, the summations in (\ref{eq:shc_so}) and (\ref{eq:bhc_so}) run over all ADOs extracted from the second level of the HEOM. This trend also extends to higher moments; namely, to evaluate $\langle I^n_{S_j}\rangle$ and $\langle I^n_{B_j}\rangle$, ADOs up to the $n$-th level of the HEOM are needed in general.

\subsection{Perturbative treatments}\label{sec:3C}

The hierarchical structure of the HEOM enables a perturbative solution to the heat currents to be systematically constructed up to a specified order in the bath coupling \cite{Trushechkin2019}. To demonstrate this, the ADOs for $|\vec{n}|+|\vec{m}|\neq0$ may first be expanded as a formal series in the dimensionless parameter $\lambda$, 
\begin{equation}\label{eq:pert_exp}
   \tilde{\rho}_{\vec{n},\vec{m}}(t) = \sum^{\infty}_{\alpha=1}\lambda^{2\alpha}\tilde{\rho}^{(\alpha)}_{\vec{n},\vec{m}}(t),
\end{equation}
where the tilde indicates a representation in the interaction picture (within Liouville space). Accordingly, this gives a perturbative expansion for the steady state current [see (\ref{eq:shc_ado}) in \ref{appen:B}];
\begin{equation}\label{eq:I_ss_pert}
	\fl \qquad \qquad I_{ss} \equiv \sum^{\infty}_{\alpha=1}\lambda^{2\alpha}I^{(\alpha)}_{ss} = -i\lim_{t\rightarrow\infty}\sum^{\infty}_{\alpha=1}\lambda^{2\alpha}\,{\rm Tr}\Big\{H_S\big[\tilde{\sigma}_z(t),\tilde{\rho}^{(\alpha)}_{\vec{0}^+_{1k},\vec{0}} + \tilde{\rho}^{(\alpha)}_{\vec{0},\vec{0}^+_{1k}}\big]\Big\}.
\end{equation}
Note that the leading-order contribution to the steady state current can be derived by restricting the HEOM to first-tier ADOs only. This is because ADOs scale as $O(\lambda^{2(|\vec{n}|+|\vec{m}|)}$) to lowest order in $\lambda$ [this can be checked by replacing $c_{jk}\rightarrow \lambda^2c_{jk}$ in (\ref{eq:ADOs})], and hence for $|\vec{n}|+|\vec{m}|\geq 2$ only contribute to (\ref{eq:I_ss_pert}) beyond second order. As such, we may determine $I^{(1)}_{ss}$ by solving the truncated first-tier equations,
\numparts
\begin{eqnarray}\label{eq:ados_tier_1}
	&\frac{d}{dt}\tilde{\rho}^{(1)}_{\vec{0}^+_{jk},\vec{0}}(t) = -iz_{jk}\tilde{\rho}^{(1)}_{\vec{0}^+_{jk},\vec{0}}(t) - ic_{jk}\tilde{\sigma}_z(t)\tilde{\rho}_S(t), \\
	&\frac{d}{dt}\tilde{\rho}^{(1)}_{\vec{0},\vec{0}^+_{jk}}(t) = iz^*_{jk}\tilde{\rho}^{(1)}_{\vec{0},\vec{0}^+_{jk}}(t) + ic^*_{jk}\tilde{\rho}_S(t)\tilde{\sigma}_z(t),
\end{eqnarray}
\endnumparts
which yields
\numparts
\begin{eqnarray}\label{eq:ados_tier_1_sol}
	\tilde{\rho}^{(1)}_{\vec{0}^+_{jk},\vec{0}}(t) &= -ic_{jk}\int^t_0d\tau\,e^{-iz_{jk}(t-\tau)}\tilde{\sigma}_z(\tau)\tilde{\rho}_S(t), \\
	\tilde{\rho}^{(1)}_{\vec{0},\vec{0}^+_{jk}}(t) &=  ic^*_{jk}\int^t_0d\tau\,e^{iz^*_{jk}(t-\tau)}\tilde{\rho}_S(t)\tilde{\sigma}_z(\tau). 
\end{eqnarray}
\endnumparts
Notice we have also replaced $\tilde{\rho}_S(\tau)\rightarrow \tilde{\rho}_S(t)$ inside each integral since this falls at the same level of approximation made in equation (\ref{eq:ados_tier_1}), i.e. $\tilde{\rho}_S(\tau) = \tilde{\rho}_S(t) + O(\lambda^2)$. By then inserting these expressions into (\ref{eq:shc_heom}), we obtain the leading term in the expansion for $I_{ss}$,
\begin{equation}\label{eq:sshc_so}
	I^{(1)}_{ss} = {\rm Tr}\{H_S\mathcal{D}^{(1)}_1[\rho^{ss}_S]\},
\end{equation}
where 
\begin{equation}\label{eq:BRME_dis}
	\mathcal{D}^{(1)}_j[\rho] = -\int^{\infty}_0d\tau\Big(C_j(\tau)[\sigma_z,\tilde{\sigma}_z(-\tau)\rho] + {\rm h.c.}\Big)
\end{equation}
is the second-order dissipator of the $j$-th bath. Equation (\ref{eq:sshc_so}) agrees with a Redfield treatment of the currents \cite{Segal2005} and describes local heat flows between the system and baths in the nonequilibrium steady state when the baths act additively. The Redfield equation for $\rho_S(t)$ can be written as \cite{Redfield1965}
\begin{equation}\label{eq:BRME}
	\frac{\partial\rho_S(t)}{\partial t} = -i[H_S,\rho_S(t)] + \sum_j\mathcal{D}^{(1)}_j[\rho_S(t)].
\end{equation}
Higher order contributions capturing non-additive effects may also be evaluated \cite{Mitchison2018}, with the corresponding fourth-order expressions given in \ref{appen:C}. 

\begin{figure}[t!]
\centering
\includegraphics[width=\textwidth]{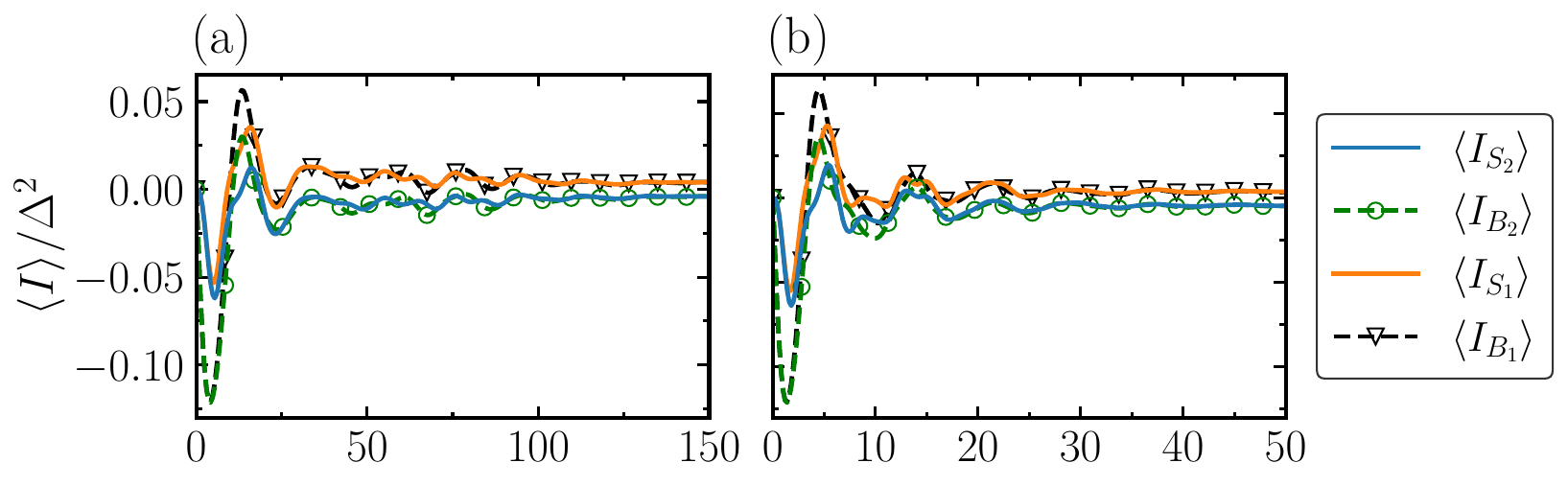}
\caption{\label{fig:2} Transient system and bath heat currents for (a) g-NESB, and (b) ug-NESB at strong system-bath coupling $\sqrt{\pi}\lambda = 0.5\Delta$ with $g=0.1\Delta$ and $g=0$, respectively. The other parameters are set to $\beta_1\Delta=0.8$, $\beta_2\Delta=1.0$, and $\Gamma_1=\Gamma_2=0.05\Delta$.}
\end{figure}

Defining the second-order noise quantity $S_{ss} \equiv \langle I^2_{S_1}\rangle$, we may construct another perturbative expansion akin to (\ref{eq:I_ss_pert}),
\begin{equation}\label{eq:noise_ss}
	   \fl S_{ss} \equiv \sum^{\infty}_{\alpha=1}\lambda^{2\alpha}S^{(\alpha)}_{ss} = \Delta^2\Big(\lambda^2C_1(0) + \lim_{t\rightarrow\infty}\sum^{\infty}_{\alpha=1}\lambda^{2\alpha}\sum_{k,k'}{\rm Tr}[\tilde{\rho}^{(\alpha)}_{\vec{0}^{+}_{1k,1k'},\vec{0}} + 2\tilde{\rho}^{(\alpha)}_{\vec{0}^+_{1k},\vec{0}^+_{1k'}} + \tilde{\rho}^{(\alpha)}_{\vec{0},\vec{0}^{+}_{1k,1k'}}]\Big),
\end{equation}
where to order $\lambda^2$,
\begin{equation}\label{eq:noise_pert}
	S^{(1)}_{ss} = \frac{\Delta^2}{\pi}\int^{\infty}_0d\omega\,J_1(\omega)\coth\left(\frac{\beta_1\omega}{2}\right). 
\end{equation}
This result is consistent with the prediction for the current noise from the fluctuation-dissipation theorem \cite{Weiss2011}; that is, $S^{(1)}_{ss}$ represents the contribution to the noise from equilibrium fluctuations. 

In the following we look to compare the predictions of the HEOM with the perturbative approach to assess the role of higher order effects (i.e. those beyond second order) on the behavior of the heat flux. 

\begin{figure}[t!]
\centering
\includegraphics[width=0.85\textwidth]{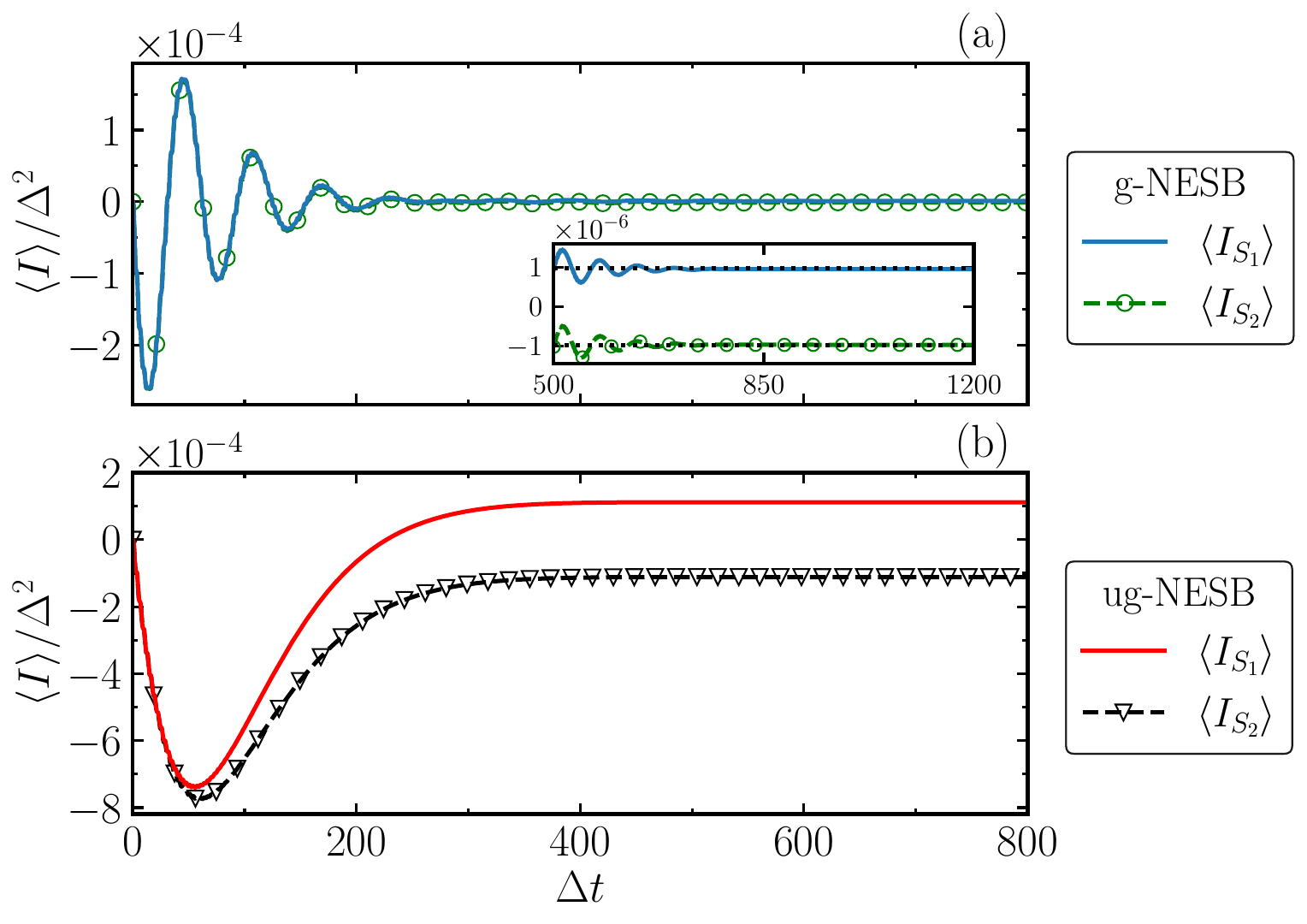}
\caption{\label{fig:3} Transient system currents for (a) the g-NESB and (b) the ug-NESB at weak system-bath coupling $\sqrt{\pi}\lambda=0.01\Delta$, where $g=0.1\Delta$ and $g=0$, respectively. The black dotted lines in the inset of panel (a) indicate the steady-state values of the corresponding currents. All other parameters are the same as in figure \ref{fig:2}.}
\end{figure}

\section{Results}\label{sec:4}

In this section we present numerical results for both the heat currents (\ref{eq:shc_heom})-(\ref{eq:bhc_heom}) and second-order moments (\ref{eq:shc_so}), (\ref{eq:noise_ss}) obtained for the g-NESB and ug-NESB. As stated above, these models correspond to when the system is coupled to the reservoirs either through a pair of interacting oscillators, or a single oscillator, respectively. For simplicity we will assume the couplings to the two baths to be equal in both cases, $\lambda_{1,2}=\lambda$, and for the system interactions to be resonant with each of the baths, $\omega_0=\Delta$ \footnote{Far from resonance, the currents in the two models are expected to exhibit the same behavior since both the gapped and ungapped spectral densities fall off as $O(1/\omega^3)$.}. All simulations are performed using the HEOM solver of the Python package QuTiP \cite{Johansson2013,Lambert2023}.

\subsection{Transient currents}

\begin{figure}[t!]
\centering
\includegraphics[width=\textwidth]{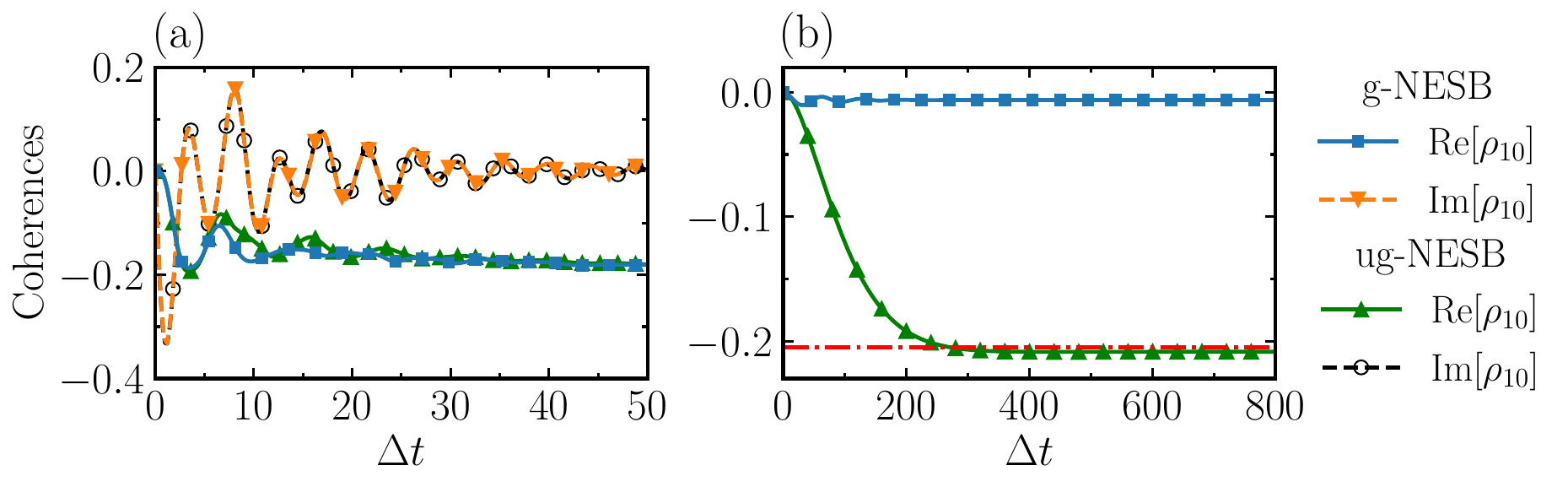}
\caption{\label{fig:4} System coherence for the g-NESB and ug-NESB at (a) strong system-bath coupling $\sqrt{\pi}\lambda=0.5\Delta$, and (b) weak coupling $\sqrt{\pi}\lambda=0.01\Delta$. ${\rm Re}[\rho_{10}]$ and ${\rm Im}[\rho_{10}]$ are the real and imaginary parts of the system coherence in the computational basis. The red dash-dotted line indicates the steady-state value of ${\rm Re}[\rho_{10}]$ in the g-NESB. All other parameters are the same as in figure \ref{fig:2}.}
\end{figure}

In figures \ref{fig:2} and \ref{fig:3}, we show plots of the transient heat flux for different values of $\lambda$, with the spin initalized in the positive $z$-direction, $\langle\sigma_z(0)\rangle=1$. Figure \ref{fig:4} shows the time evolution of the system coherences for the same set of bath parameters. At strong coupling $\sqrt{\pi}\lambda=0.5\Delta$, we see that the energy currents for the g-NESB and ug-NESB show qualitatively similar behavior; in particular, the finite contribution of the interaction energy at these coupling strengths plays a similar role in the heat current dynamics for the two cases. The rate at which the heat currents reach their stationary values, however, is generally slower for the baths with gapped spectra. This behaviour is also reflected in the system coherences which undergo relaxation on a similar time scale as the heat currents. 

Conversely, as the coupling strength is further decreased to $\sqrt{\pi}\lambda=0.01\Delta$, transient characteristics of the heat flux begin to deviate for the different bath configurations; see figure \ref{fig:3}. Here we only plot the system heat currents since the interaction energies are negligible for all times. Interestingly, we now observe a decoupling between time scales determining the relaxation of the heat currents and the spin coherences for the g-NESB---i.e. the coherences relax much more slowly towards their steady-state than the corresponding current. This is exemplified in figure \ref{fig:4}(b) where the real part of the off-diagonal elements of $\rho_S(t)$ is displayed alongside it's asymptotic value (red dash-dotted line). Additionally, at weak coupling, the gapped baths induce a much smaller rate of energy transfer in the long time limit, contrasting with the ungapped case in which the HEOM and Redfield results are in good agreement (not shown). 

\begin{figure}[b!]
\centering
\includegraphics[width=\textwidth]{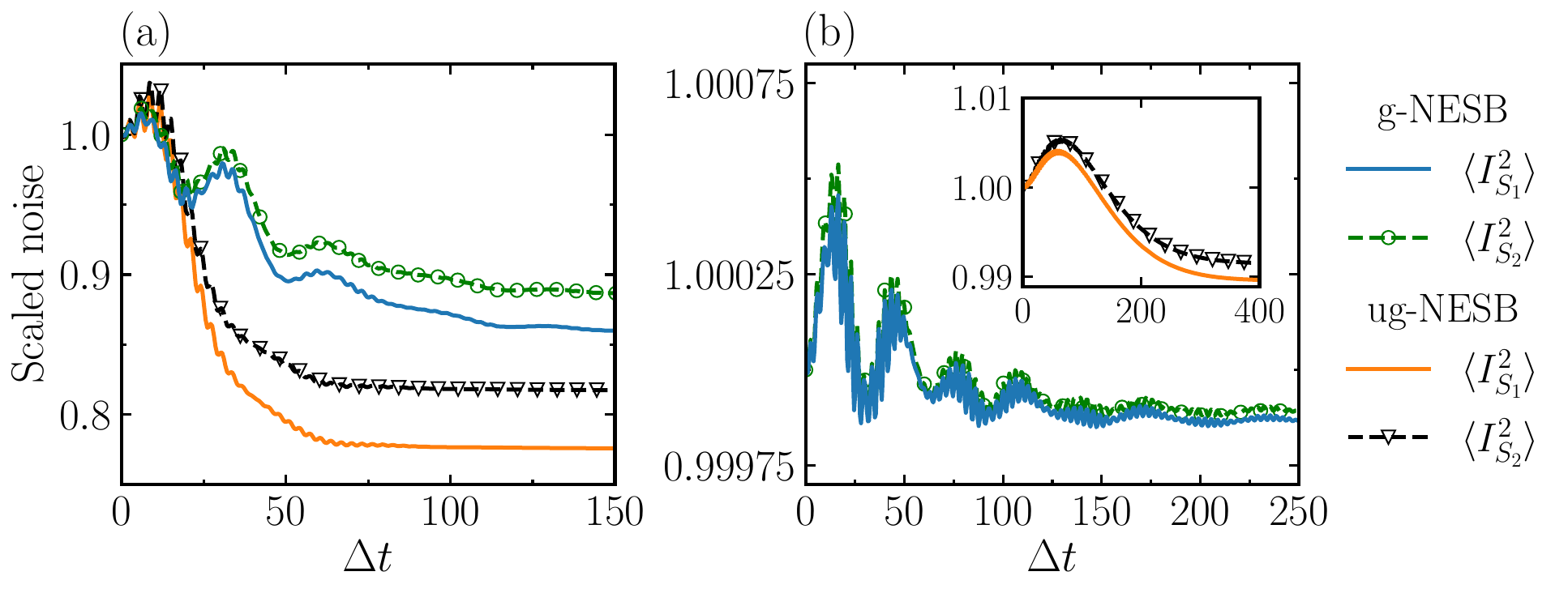}
\caption{\label{fig:5} Transient noise (\ref{eq:shc_so}) for the g-NESB and ug-NESB at (a) moderate (strong) $\sqrt{\pi}\lambda = 0.1\Delta$, and (b) weak coupling $\sqrt{\pi}\lambda = 0.01\Delta$. The inset shows the ug-NESB currents on a longer time scale. All other parameters are the same as in figure \ref{fig:2}.}
\end{figure}

Figure \ref{fig:5} displays the scaled dynamical quantity $\langle I^2_{S_j}\rangle/C_j(0)$ characterizing the strength of out-of-equilibrium fluctuations relative to those in equilibrium. Note that our analysis here is restricted to second moments of $I_{S_j}$ only, since bath moments beyond $\langle I_{B_j}\rangle$ diverge for the spectral densities under consideration. At moderate (strong) system-bath coupling, the scaled noise for the two models exhibits similar short time behavior, but reaches its steady-state value on different time scales. The difference in relaxation times becomes more substantial in the weak coupling regime, although in contrast with the heat currents in figure \ref{fig:3}, the noise relaxes faster for the g-NESB than for the ug-NESB. The transient noise for the g-NESB additionally shows quantum beats which are not present for the ug-NESB. 

\begin{figure}[t!]
\centering
\includegraphics[width=\textwidth]{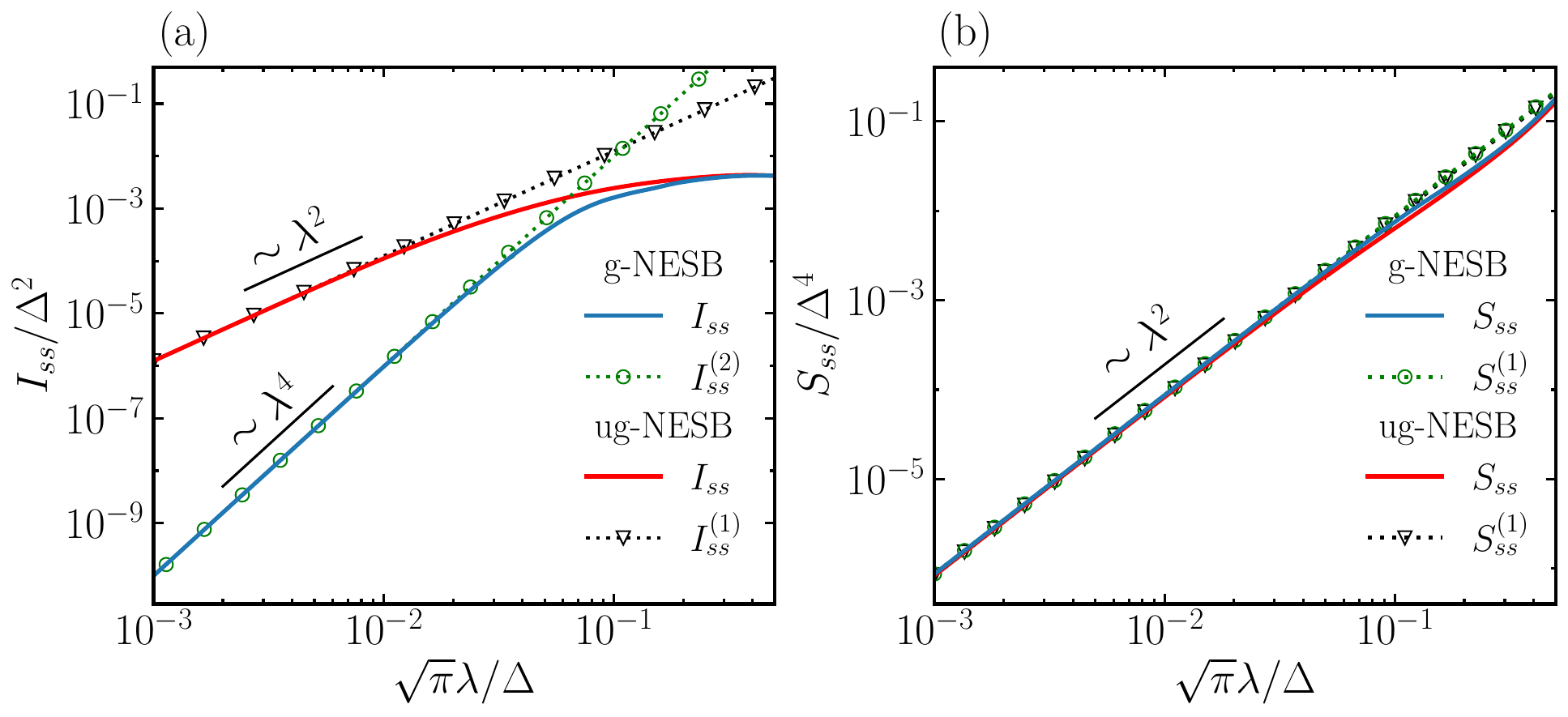}
\caption{\label{fig:6} Stationary (a) heat current and (b) noise $S_{ss} = \langle I^2_{S_1}\rangle_{ss}$ as a function of the coupling strength $\lambda$ for the g-NESB and ug-NESB. In (a), the green open circles and black open triangles represent perturbative results for the steady state current obtained from (\ref{eq:I_ss_pert}). The other bath parameters are the same as in figure \ref{fig:2}.}
\end{figure}

\subsection{Steady state current}

To gain further insight into the differences in the behavior of the heat currents for the two setups, in figure \ref{fig:6}(a) we show the stationary heat currents (\ref{eq:I_ss}) as a function of the coupling strength. The solid lines represent the `exact' results obtained from the converged HEOM, while open points and dashed lines represent the perturbative results of the truncated HEOM approaches. The acronyms HEOM-1 and HEOM-2 reference the HEOM result evaluated up to second and fourth order in the coupling---i.e. $I^{(1)}_{ss}$ and $I^{(2)}_{ss}$---respectively.

In the weak and ultra-weak coupling regimes, the scaling of the stationary current with $\lambda$ is noticeably different for the g-NESB and ug-NESB. In the latter case, the current scales quadratically in the coupling strength, $I_{ss}\sim\lambda^2$, and thus agrees with the HEOM-1 results which treat the heat current up to second-order in the coupling. On the other hand, for baths with gapped spectra, the current scales as $I_{ss}\sim \lambda^4$. The HEOM-1 therefore cannot capture the steady-state behavior of the current for this case, which must be treated to at least fourth-order in $\lambda$. In particular, a second-order approach predicts no heat transfer here at all, since the dissipator (\ref{eq:BRME_dis}) vanishes at the system Bohr frequencies, i.e $J_{j}(\pm\Delta)=0$ and $J_j(0)=0$.

At strong system-bath coupling, the energy currents exhibit a turnover behavior around $\sqrt{\pi}\lambda=0.1\Delta$, at which point the perturbative approaches no longer agree with exact results. Deeper into the turnover region the two solid lines also converge, indicating that spin transitions mediating the heat flow become largely insensitive to the gap in the bath spectra when $\lambda$ becomes large. 

Next, we display the steady state noise $S_{ss}$ in figure \ref{fig:6}(b) plotted against the coupling strength $\lambda$ for the two bath configurations. It can be seen that the g-NESB and ug-NESB display similar noise characteristics in contrast with the different scaling profiles observed for the heat currents. In particular, at weak system-bath coupling the noise scales quadratically with $\lambda$ in both cases as predicted by the fluctuation-dissipation theorem. Furthermore, the noise unlike the current $I_{ss}$ does not exhibit turnover behavior in either of the two models at strong coupling. As such, the perturbative and exact expressions for $S_{ss}$ also adopt a similar functional form in the turnover region of the current: discrepancies only occur between approaches for coupling strengths above $\lambda\approx 0.1\Delta$. 

\begin{figure}[t!]
\centering
\includegraphics[width=\textwidth]{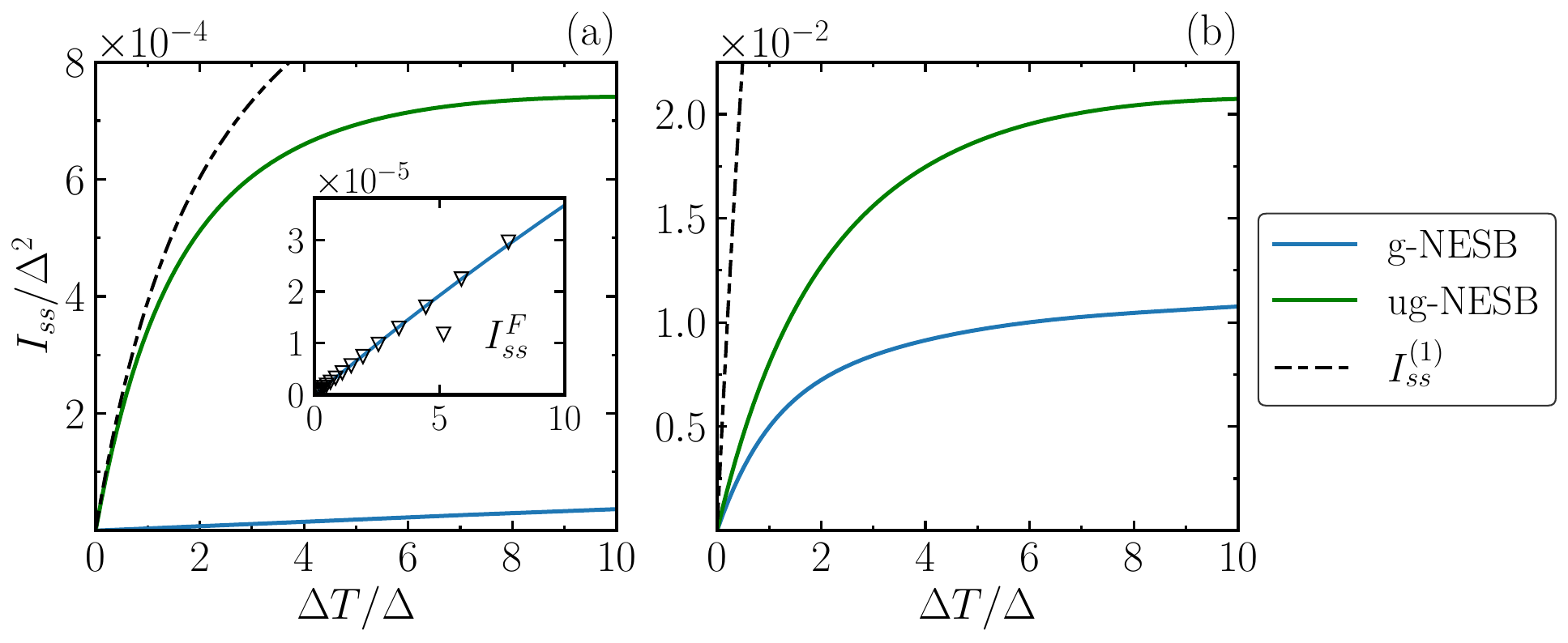}
\caption{\label{fig:7} Stationary heat current as a function of the temperature difference $\Delta T=T_1-T_2$ for (a) weak coupling $\sqrt{\pi}\lambda=0.01\Delta$, and (b) moderate (strong) coupling $\sqrt{\pi}\lambda=0.1\Delta$, with $T_2=\Delta$. The solid blue and green lines correspond to the exact HEOM results for $g=0.1\Delta$ and $g=0$, while the black dashed line represents the Redfield results. The inset shows the result of the simulated g-NESB against Fourier's law $I^F_{\rm ss}=\kappa\Delta T$. All other parameters are the same as in figure \ref{fig:6}.}
\end{figure}

Lastly, figure \ref{fig:7} illustrates how the stationary currents for each model depends on the temperature bias $\Delta T = T_1-T_2$, assuming the temperature of the cold bath to be fixed to $T_2 = \Delta$. We compare the exact HEOM simulations with those obtained from the second-order Redfield equation (\ref{eq:BRME_dis}) for moderate (strong) and weak system-bath coupling. When the baths are close to equilibrium $\Delta T\ll1$, we find the current scales linearly with $\Delta T$ for moderate and weak coupling as would be expected from Fourier's law, $I^{F}_{ss}=\kappa\Delta T$, where
\begin{equation}
	\kappa = \lim_{T_1\rightarrow T_2}\,\frac{I_{ss}(T_1)}{T_1-T_2}
\end{equation}
is the thermal conductance. As the temperature difference grows in magnitude, however, the two models display differing behavior. Within the ug-NESB, the heat current saturates as $\Delta T$ is tuned away from the linear response regime. This saturation also occurs in the g-NESB but the current approaches it's asymtoptic value at a slower rate, thereby causing Fourier's law to hold over a wider temperature range than for the ug-NESB, particularly at weaker coupling. We note further that while the Redfield equation (\ref{eq:sshc_so}) predicts a similar turnover as the HEOM, it tends to substantially overestimate the heat flow in the ug-NESB.

\section{Summary}\label{sec:5}

We have analyzed heat transport in a structured-environment version of the NESB, where energy exchange between the system and Ohmic reservoirs is mediated by pairs of coupled harmonic oscillators. Compared to the case of a single mediating oscillator considered in \cite{Aurell2021,Yamamoto2021}, the transport properties are found to be significantly altered in both the transient and steady-state regimes (with the exception of the steady-state current noise), specifically at weak system-bath coupling. In particular, the steady-state heat current displays a non-trivial scaling that cannot be captured using standard perturbative techniques including the second-order master equation (\ref{eq:BRME}). To rigorously treat higher-than-second-order effects we have used the numerically exact HEOM, which facilitates the calculation of arbitrary moments of system and bath current operators in terms of ADOs. For example, averages depend on ADOs extracted from the first-tier of the HEOM \cite{Zhu2012}. The calculation of higher moments has also been achieved through extending the generating functional technique used by Kato and Tanimura \cite{Kato2016} to moments beyond first-order on par with the path integral treatment of \cite{Song2017}. It should be further noted that our implementation of the HEOM is distinct from that developed by Cerrillo \etal \cite{Cerrillo2016}, which is based on the full counting statistics formalism \cite{Esposito2009}.

Overall, our results highlight the possibility of using reservoir engineering to non-trivially influence the properties of thermal transport through a quantum system. In this aspect, practical applications may be found in the design of nanoscale heat devices. In particular, analogous models investigating heat transport through superconducting circuits, where a similar resonator-qubit-resonator architecture is considered, indicate the potential for our model to be implemented on such a platform.

Future work could potentially examine the effect of asymmetry on the transport properties of the model, including rectification of the currents that arises with noncommutative system-bath coupling operators \cite{AntoSztrikacs2022,Duan2020} and-or asymmetric coupling strengths \cite{Segal2005,Ruokola2009}, and in particular, whether the diode effect is enhanced or diminished for environments possessing gapped spectra. Alternatively, generalisations of the model to include couplings between components of the different baths in the spirit of \cite{Xu2021,Reuther2010} could be analyzed. Finally, applications to continuous thermal machines, such as autonomous refrigerators and heat engines, may also be of interest.

\ack

We gratefully acknowledge Camille Lombard Latune and Erik Aurell for helpful discussions and feedback on the manuscript. We also thank the Centre for High Performance Computing (CHPC) of South Africa for generous allocation of computing time. This work is supported by funding from the South African Quantum Technology Initiative (SA QuTI) through the Department of Science and Innovation of South Africa and the National Research Foundation (Grant number UID:65212).

\section*{References}

\bibliography{main_v2}

\appendix 

\section{Derivation of the effective spectral densities}\label{appen:A}

Here, we derive an analytical expression for the effective bath spectral densities (\ref{eq:J_eff}) by comparing the system dynamics within the two bath configurations. The derivation follows reference \cite{Nazir2018} and is based on the equivalent representation of the Hamiltonian (\ref{eq:H_B}) in terms of normal modes. We shall also like \cite{Nazir2018} limit ourselves to the case of just a single bath, since the generalization to multiple independent baths is straightforward. 

\subsection{Original representation}

We begin by writing down the Heisenberg equations of motion of the system and bath in the original representation [see (\ref{eq:H})]. Denoting by $O_S$ a generic system operator, the equations read ($i=1,2$)
\begin{eqnarray*}
	\fl \qquad \qquad \frac{d}{dt}O_S(t) &= -iA_S(t) -i\lambda V_S(t)X_2(t),  \\
	\fl \qquad \qquad \frac{d}{dt}a_i(t) &= -i\omega_0a_i(t) - igX_{\bar{i}}(t) - i2\delta\omega_iX_i(t) - i\sum_k\tilde{g}_{k_i}X_{k_i}(t) - i\lambda V(t)\delta_{i2},  \\
	\fl \qquad \qquad \frac{d}{dt}\tilde{b}_{k_i}(t) &= -i\tilde{\omega}_k\tilde{b}_{k_i}(t) - i\tilde{g}_{k_i}X_i(t),
\end{eqnarray*}
where $A_S = [O_S,H_S]$ and $V_S=[O_S,V]$, $V=V^{\dagger}$ is an arbitrary system observable, $\delta\omega_i = 1/\pi\int^{\infty}_0d\omega\,\tilde{J}_i(\omega)/\omega$ is the energy shift of the $i$-th oscillator due to renormalization, and the bar notation $\bar{i}$ denotes the conjugate index to $i$, i.e. $\bar{1}=2$. The derivation proceeds by eliminating the bath variables from the equation of motion for $O_S(t)$. Hence, we Fourier-transform the above equations according to $\widehat{O}(\nu) = \int^{\infty}_{-\infty}dt\,O(t)e^{-i\nu t}$, which yields 
\begin{eqnarray}\label{eq:HeisenbergFT}
	\fl \qquad \qquad \nu\widehat{O}_S(\nu) &= \widehat{A}_S(\nu) + \lambda\int d\nu'\,\widehat{V}_S(\nu')\widehat{X}_2(\nu-\nu'), \\
	\fl \qquad \qquad \nu\widehat{a}_i(\nu) &= \omega_0\widehat{a}_i(\nu) + g\widehat{X}_{\bar{i}}(\nu) + 2\delta\omega_i\widehat{X}_i(\nu) + \sum_k\tilde{g}_{k_i}\widehat{X}_{k_i}(\nu) + \lambda\widehat{V}(\nu)\delta_{i2}, \nonumber\\
	\fl \qquad \qquad \nu\widehat{\tilde{b}}_{k_i}(\nu) &= \tilde{\omega}_{k_i}\widehat{\tilde{b}}_{k_i}(\nu) + \tilde{g}_{k_i}\widehat{X}_i(\nu). \nonumber
\end{eqnarray}
Care must be taken when deriving similar expressions for operators such as $\widehat{a}^{\dagger}_i(\nu)$, since the Fourier transform of $a^{\dagger}_i(t)$ is in general \textit{not} the same as the adjoint of $a_i(t)$ in the Fourier domain, i.e. $\widehat{a}^{\dagger}_i(\nu) \neq [\widehat{a}_i(\nu)]^{\dagger}$ \cite{Nazir2018} (the dagger is kept on Fourier transformed variables to simplify the notation). In this way, by solving for $\widehat{\tilde{b}}_{k_i}(\nu)$ and $\widehat{\tilde{b}}^{\dagger}_{k_i}(\nu)$,
\begin{eqnarray*}
	\widehat{\tilde{b}}_{k_i}(\nu) &= \frac{\tilde{g}_{k_i}}{z-\omega_k}\widehat{X}_i(\nu), \\
	\widehat{\tilde{b}}^{\dagger}_{k_i}(\nu) &= \frac{-\tilde{g}_{k_i}}{z+\omega_k}\widehat{X}_i(\nu),
\end{eqnarray*}
and substituting these into the second equality of (\ref{eq:HeisenbergFT}), we obtain the following expression for $\widehat{X}_2(\nu)$:
\begin{equation}\label{eq:X_2}
	\fl \, \widehat{X}_2(\nu) = \frac{2\omega_0\lambda}{\nu^2-\omega^2_0}\left[1 - \frac{4\omega_0}{\nu^2-\omega^2_0}\left[\delta\omega_{c_2} + \frac{1}{\pi}\int^{\infty}_0d\omega\frac{\tilde{J}_i(\omega)\omega}{\nu^2-\omega^2}\right] - \frac{4\omega^2_0g^2}{(\nu^2 - \omega^2_0)^2}\right]^{-1}\widehat{V}(\nu).
\end{equation}
For simplicity, we have assumed the first oscillator to be fully decoupled from its local reservoir, i.e., $\tilde{g}_{k_1}=0$, $\forall k$. We also assume the spectral density characterizing the interaction between the second oscillator and its reservoir to be Ohmic with Lorentz-Drude cutoff ($\gamma_1=0$),  
\begin{equation}
	\tilde{J}_i(\omega) = \frac{2\gamma_i\omega\omega^2_c}{\omega^2+\omega^2_c}.
\end{equation}
Taking the continuum limit and substituting (\ref{eq:X_2}) into the first equality of (\ref{eq:HeisenbergFT}) then gives
\begin{equation}\label{eq:kernel_eq}
	\nu\widehat{O}_S(\nu) = \widehat{A}_S(\nu) + \int d\nu'\,\widehat{V}_S(\nu')K(\nu-\nu')\widehat{V}(\nu-\nu')
\end{equation}
where
\begin{equation}
	K(\nu) = \frac{-2\omega_0\lambda^2}{\nu^2 - \omega^2_0 - 4\omega_0W_2(\nu) - \frac{4\omega^2_0g^2}{\nu^2-\omega^2_0}}, 
\end{equation}
and 
\begin{equation}\label{eq:W_2}
	W_2(\nu) = \delta\omega_{c_2} + \frac{1}{\pi}\int^{\infty}_0d\omega\frac{\tilde{J}_2(\omega)\omega}{\nu^2-\omega^2} =  \delta\omega_{c_2} + \frac{1}{2\pi}\int^{\infty}_{-\infty}d\omega\frac{\tilde{J}_2(\omega)}{\nu-\omega}.
\end{equation}
The quantity $W_2(\nu)$ may now be evaluated using the residue theorem, given that $\tilde{J}_2(\omega)$ has poles located at $i\omega_c$ ($-i\omega_c$) in the upper (lower) half complex plane. In doing so, we obtain
\begin{equation}
	W_2(\nu) = \frac{-i\gamma_2\nu}{1 + (\nu/\omega_c)^2} + \gamma_2\omega_c - \frac{\gamma_2\omega_c}{1 + (\nu/\omega_c)^2},
\end{equation}
where taking $\omega_c$ to infinity yields $\lim_{\omega_c\rightarrow\infty}W_2(\nu) = -i\gamma_2\nu$. Thus, (\ref{eq:W_2}) simplifies to
\begin{equation}\label{eq:kernel}
	K(\nu) =  \frac{-2\omega_0\lambda^2}{\nu^2 - \omega^2_0 + i4\gamma_2\omega_0\nu - \frac{4\omega^2_0g^2}{\nu^2-\omega^2_0}}, \qquad \omega_c\rightarrow\infty.
\end{equation}

\subsection{Non-interacting representation}

Our next step is to derive an analogous expression for the kernel $K(z)$ in the non-interacting bath representation. Following the same procedure as before, we write down the Heisenberg equations of motion of the full system [see (\ref{eq:H_B_eff})-(\ref{eq:H_I_eff})],
\begin{eqnarray*}
	\frac{d}{dt}O_S(t) &= -iA_S(t) -iV_S(t)\sum_kg_k[b_k(t) + b^{\dagger}_k(t)], \\
	\frac{d}{dt}b_k(t) &= -i\omega_kb_k(t) -ig_kV(t), 
\end{eqnarray*}
and eliminate the bath variables from the first equality by transforming these equations into the Fourier domain. This recovers the same relation from (\ref{eq:kernel_eq}), but with a different kernel
\begin{equation}\label{eq:kernel_nonint}
	K(\nu) = \frac{2}{\pi}\int^{\infty}_0d\omega\frac{J(\omega)\omega}{\nu^2-\omega^2} = \frac{1}{\pi}\int^{\infty}_{-\infty}d\omega\frac{J(\omega)}{\nu-\omega}
\end{equation}
written in terms of the effective bath spectral density $J(\omega)=-J(-\omega)$. Note that the kernels derived in (\ref{eq:kernel}) and (\ref{eq:kernel_nonint}) are formally equivalent since the correspondence between the two bath representations is exact. Therefore, under analytic continuation of $K(\nu)$ to the lower half complex plane, $\nu=\omega-i\varepsilon$, the spectral density $J(\omega$) is determined as
\[
	\fl \qquad J(\omega) = \lim_{\varepsilon\rightarrow0^+}{\rm Im}\left[K(\omega-i\varepsilon)\right] = \frac{2\lambda^2\omega_0(\omega^2-\omega^2_0)^2\Gamma\omega}{[(\omega^2-\omega^2_0)^2-4\omega^2_0g^2]^2+(\omega^2-\omega^2_0)^2\Gamma^2\omega^2},
\]
with $\Gamma = 4\gamma_2\omega_0$.

\section{Quantum heat currents in terms of ADOs}\label{appen:B}

In this appendix we outline the derivation of the heat currents following a generating functional approach. The approach we implement has previously been employed in \cite{Kato2016} to derive analogous expressions for averages of system and bath heat currents within the HEOM framework. Here, we extend their formulation to also include the derivation of higher-order moments in accordance with \cite{Song2017,Zhu2012}.

\subsection{Generating functional}

To fix the notation, we first note that the total density matrix in the interaction picture evolves according to
\begin{equation}\label{eq:rho_tot}
\tilde{\rho}_{\rm tot}(t) = \mathcal{T}\{e^{-i\sum_j\int^t_0ds\,\tilde{\mathcal{L}}_{I_j}(s)}\}\rho_{\rm tot}(0),
\end{equation}
where $\mathcal{T}$ is the chronological time ordering operator \cite{Breuer2002}, $\tilde{\mathcal{L}}_{I_j}(t)\rho=[\tilde{H}_{I_j}(t),\rho]$ is the Liouville superoperator associated with the interaction Hamiltonian $\tilde{H}_{I_j}(t)=\tilde{V}_j(t)\tilde{B}_j(t)$ [see (\ref{eq:H_I_eff})], and $V_j=V^{\dagger}_j$ an arbitrary system operator. We recall that the tilde indicates a time dependence within the interaction picture, $\tilde{O}(t)=e^{i(H_S+H_B)t}Oe^{-i(H_S+H_B)t}$. Moreover, by defining the left and right acting superoperators,
\begin{equation}
 	O^{+}\rho\equiv O\rho, \qquad O^{-}\rho\equiv\rho O,
\end{equation}
the Liouvillian may equivalently be represented as 
\begin{equation}
\tilde{\mathcal{L}}_{I_j}(t) = \tilde{H}_{I_j}(t)^+ - \tilde{H}_{I_j}(t)^-.
\end{equation}
We now modify each $\tilde{\mathcal{L}}_{I_j}(t)$ to include an additional time-dependent source term $\phi_j$,
\begin{equation}\label{eq:counting_field}
 \tilde{\mathcal{L}}_{I_j}[\phi_j](t) =  \tilde{\mathcal{L}}_{I_j}(t) + \phi_j(t)\tilde{\mathcal{O}}_j(t),
\end{equation}
where $\tilde{\mathcal{O}}_j(t)$ represents an arbitrary superoperator (system or bath). As such, one can define the following moment generating functional 
\begin{eqnarray}\label{eq:MGF}
	\tilde{\rho}^{\{\mathcal{O}\}}_S(\phi,t) &= {\rm Tr}_B[\tilde{\rho}_{\rm tot}(\phi,t)]  \nonumber\\
				      			           &= \bigg\langle\mathcal{T}\prod_j\exp\bigg[-i\int^t_0ds\,\tilde{\mathcal{L}}_{I_j}[\phi_j](s)\bigg]\bigg\rangle_B\rho_S(0),
\end{eqnarray}
with $\langle\cdot\rangle_B\equiv \prod_j{\rm Tr}_{B_j}[\cdot \,e^{-\beta_jH_{B_j}}]/Z_{B_j}$. Note that it is possible to factorize (\ref{eq:rho_tot}) into a product of exponentials because each of the integrals commute under time ordering \cite{Ziman1969}. 

The generating functional $\tilde{\rho}^{\{\mathcal{O}\}}_S(\phi,t)$ enables arbitrary moments of $\tilde{\mathcal{O}}_j(t)$ to be extracted via the general rule
\begin{equation}\label{eq:moments}
	\fl \qquad \Big\langle\mathcal{T}\tilde{\mathcal{O}}_{j_n}(t_n)...\tilde{\mathcal{O}}_{j_1}(t_1)\prod_j e^{-i\int^t_0ds\,\tilde{\mathcal{L}}_{I_j}(s)}\rho_S(0)\Big\rangle_B = i^n\frac{\delta^n\tilde{\rho}^{\{\mathcal{O}\}}_S(\phi,t)}{\delta \phi_{j_1}(t_1)...\delta \phi_{j_n}(t_n)}\bigg|_{\phi=0}.
\end{equation}
This will be used to derive explicit expressions for the system and bath heat currents $I_{S_j}$, $I_{B_j}$, and their corresponding moments, as explained below. 

\subsection{System heat currents}

Starting with the system heat current (\ref{eq:shc}), we first set $\tilde{\mathcal{O}}_j(t) = \tilde{B}_j(t)^+$ and use (\ref{eq:moments}) to write 
\begin{equation}\label{eq:shc_func_der}
	\langle I_{S_j}\rangle = {\rm Tr}_S\left[\tilde{S}_j(t)\frac{\delta}{\delta\phi_j(t)}\tilde{\rho}^{B}_S(\phi,t)|_{\phi=0}\right],
\end{equation}
where
\begin{equation}
	\tilde{S}_j(t) = [\tilde{H}_S(t),\tilde{V}_j(t)].
\end{equation}
The evaluation of the functional derivative above requires us to initially obtain a closed form expression for $\tilde{\rho}^{B}_S(\phi,t)$. To this end, we note that since each of the baths obeys Gaussian statistics, the righthand side of (\ref{eq:MGF}) can be evaluated using Wick's theorem,
\begin{eqnarray}
	 &\bigg\langle\mathcal{T}\prod_j\exp\bigg[-i\int^t_0ds\,\tilde{\mathcal{L}}_{I_j}[\phi_j](s)\bigg]\bigg\rangle_B \nonumber\\
	 &\qquad = \mathcal{T}\prod_j\exp\bigg[-\frac{1}{2}\int^t_0ds\int^t_0du\,\mathcal{T}\langle\tilde{\mathcal{L}}_{I_j}[\phi_j](s)\tilde{\mathcal{L}}_{I_j}[\phi_j](u)\rangle_{B_j}\bigg].
\end{eqnarray}
Here, we've used the idempotency of the time ordering superoperator, $\mathcal{T}^2=\mathcal{T}$, as well as that $\langle\prod_j(\cdot)\rangle_B=\prod_j\langle\cdot\rangle_{B_j}$, and $\langle\tilde{\mathcal{L}}_{I_j}(t)\rangle_{B_j}=0$. By now using
\begin{eqnarray}
&\mathcal{T}\prod_j\exp\bigg[-\frac{1}{2}\int^t_0ds\int^t_0du\,\mathcal{T}\langle\tilde{\mathcal{L}}_{I_j}[\phi_j](s)\tilde{\mathcal{L}}_{I_j}[\phi_j](u)\rangle_{B_j}\bigg] \nonumber\\
&\qquad = \mathcal{T}\prod_j\exp\bigg[-\int^t_0ds\int^s_0du\,\langle\tilde{\mathcal{L}}_{I_j}[\phi_j](s)\tilde{\mathcal{L}}_{I_j}[\phi_j](u)\rangle_{B_j}\bigg],
\end{eqnarray}
and evaluating the partial trace over the baths [see (\ref{eq:counting_field})], we get
\begin{eqnarray}
	\fl \tilde{\rho}^{B}_S(\phi,t) &= \mathcal{T}\bigg\{\prod_j\exp\bigg[-\int^t_0ds\int^s_0du\Big(\phi_j(u)C_j(s-u)\tilde{V}_j(s)^- + \phi_j(s)\tilde{\mathcal{B}}_j(s,u)\nonumber \\
	\fl &\qquad\qquad\quad\,\, + \phi_j(s)\phi_j(u)C_j(s-u)\Big) - \int^t_0ds\,\tilde{\mathcal{W}}_j(s)\bigg]\bigg\}\rho_S(0), \\
	\fl &\equiv \mathcal{T}\bigg\{\prod_j\exp\bigg[-\int^t_0ds\,\tilde{\mathcal{W}}_j[\phi_j](s)\bigg]\bigg\}\rho_S(0), \nonumber 
\end{eqnarray}
where we've introduced the influence phase superoperators $\tilde{\mathcal{W}}_j(s)\equiv \tilde{\mathcal{W}}_j[\phi_j=0](s)$, and
\begin{equation}
\tilde{\mathcal{B}}_j(s,u) = C_j(s-u)\tilde{V}_j(u)^+ - C^*_j(s-u)\tilde{V}_j(u)^-. 
\end{equation}
Having obtained an exact expression for $\tilde{\rho}^B_S(\phi,t)$, one can next compute it's functional derivative according to \cite{Parr1989}
\begin{eqnarray}
	 \fl \frac{d}{d\varepsilon}\tilde{\rho}^B_S(\phi_j+\varepsilon\eta_j,t)|_{\varepsilon=0} &= 
									      -\int^t_0ds\,\mathcal{T}\bigg\{\int^s_0du\Big[C_j(s-u)\Big(\eta_j(u)\tilde{V}_j(s)^- + \eta_j(u)\phi_j(s)+\eta_j(s)\phi_j(u)\Big) \nonumber\\
									      &\qquad +\eta_j(s)\tilde{\mathcal{B}}_j(s,u)\Big]\tilde{\mathcal{F}}[\phi]\bigg\}\rho_S(0) \label{eq:func_derv_B} \\
									      & \equiv \int^t_0ds\,\eta_j(s)\frac{\delta}{\delta\phi_j(s)}\tilde{\rho}^{B}_S(\phi,s). \nonumber
\end{eqnarray}
Here, $\eta_j(t)$ is a suitable test function, and $\tilde{\mathcal{F}}[\phi]$ is the Feynman-Vernon influence functional incorporating the source terms $\phi$, 
\begin{equation}
	\tilde{\mathcal{F}}[\phi] = \prod_j\exp\bigg[-\int^t_0ds\,\tilde{\mathcal{W}}_j[\phi_j](s)\bigg].
\end{equation}

Let us denote by $\mathcal{R}=\{(s,u):0\leq s \leq t, \, 0\leq u\leq s\}$ the triangular region of integration in (\ref{eq:func_derv_B}). Over this region the integrals containing the terms $\eta_j(u)$ may be rewritten as
\begin{eqnarray*}
 	\int\int_{\mathcal{R}}ds\,du\,\eta_j(u)C_j(s-u)\tilde{V}_j(s)^- &= \int^t_0ds\int^t_sdu\,\eta_j(s)C_j(u-s)\tilde{V}_j(u)^-, \\
	\int\int_{\mathcal{R}}ds\,du\,\eta_j(u)C_j(s-u)\phi_j(s) &= \int^t_0ds\int^t_sdu\,\eta_j(s)C_j(u-s)\phi_j(u),
\end{eqnarray*}
so that by substituting these expressions back into (\ref{eq:func_derv_B}), we can directly read off the functional derivative of $\tilde{\rho}^{B}_S(\phi,t)$. In doing so, we get 
\begin{eqnarray}\label{eq:MGF_fd}
	\fl \frac{\delta}{\delta\phi_j(t)}\tilde{\rho}^{B}_S(\phi,t) &= -\mathcal{T}\bigg\{\bigg[\int^t_0ds\,\tilde{\mathcal{B}}_j(t,s) + \lim_{s\rightarrow t}\int^t_sdu\,C_j(u-s)\big[\tilde{V}_j(u)^- + \phi_j(u)\big] \nonumber\\
	&\qquad\qquad + \int^t_0ds\,C_j(t-s)\phi_j(u)\bigg]\tilde{\mathcal{F}}(t)\bigg\}\rho_S(0) \nonumber\\
	&= -\mathcal{T}\bigg\{\bigg[\int^t_0ds\,\tilde{\mathcal{B}}_j(t,s) + \int^t_0ds\,C_j(t-s)\phi_j(u)\bigg]\tilde{\mathcal{F}}(t)\bigg\}\rho_S(0),
\end{eqnarray}	
which in turn gives us the following expression for the system heat current,
\begin{eqnarray}
	\fl \langle I_{S_j}\rangle &= -{\rm Tr}_S\left[\int^t_0ds\,\mathcal{T}\{\tilde{S}_j(t)\tilde{\mathcal{B}}_j(t,s)\tilde{\mathcal{F}}(t)\}\rho_S(0)\right] \nonumber\\		     
		      \fl &\equiv -\int^t_0ds\,C_j(t-s)\left\langle\mathcal{T}\tilde{S}_j(t)^+\tilde{V}_j(s)^+\tilde{\mathcal{F}}(t)\right\rangle_S + \int^t_0ds\,C^*_j(t-s)\left\langle\mathcal{T}\tilde{S}_j(t)^+\tilde{V}_j(s)^-\tilde{\mathcal{F}}(t)\right\rangle_S. \nonumber\label{eq:shc_scf}
\end{eqnarray}
We can now insert the exponential decomposition of the bath correlation functions (\ref{eq:corr_func}) into the above to obtain
\begin{equation}\label{eq:shc_ado}
        \langle I_{S_j}\rangle = -i\sum_k{\rm Tr}[S_j(\hat{\rho}_{\vec{0}^+_{jk},\vec{0}} + \hat{\rho}_{\vec{0},\vec{0}^+_{jk}})].
\end{equation}

The derivation of all higher order currents can be achieved through evaluating derivatives of the moment generating functional (\ref{eq:MGF}) for $n>1$. In particular, the $n$-th order moment of the system heat current is determined from
\begin{equation}
	\langle I^n_{S_j}\rangle = {\rm Tr}\left[\tilde{S}^n_j(t)\frac{\delta^n}{\delta\phi_j(t)^n}\tilde{\rho}^{B}_S(\phi,t)|_{\phi=0}\right]. \label{eq:moments_fd}
\end{equation}
We now proceed to evaluate $\langle I^n_{B_j}\rangle$ systematically starting with the $n=2$ case. In this regard, we note that the second derivative of $\tilde{\rho}_S(\phi,t)$ may be formally written as [see (\ref{eq:MGF_fd})]
\begin{equation}
	\frac{\delta^2\tilde{\rho}^{B}_S(\phi,t)}{\delta\phi_j(t)^2}= -\mathcal{T}\bigg\{\frac{\delta\Lambda_j[\phi_j]}{\delta\phi_j(t)}\tilde{\mathcal{F}}[\phi] + \Lambda_j[\phi_j]\frac{\delta\tilde{\mathcal{F}}[\phi]}{\delta\phi_j(t)}\bigg\}\rho_S(0), 
\end{equation}
where 
\begin{eqnarray}
	    \Lambda_j[\phi_j] = \int^t_0ds\big[\tilde{\mathcal{B}}_j(t,s) + C_j(t-s)\phi_j(s)\big]. 
\end{eqnarray}
It is straightforward to verify that
\begin{equation}
	\frac{\delta\Lambda_j[\phi_j]}{\delta\phi_j(t)} = \frac{\delta}{\delta\phi_j(t)}\int^t_0ds\,C_j(t-s)\phi_j(s) = C_j(0),
\end{equation}
with $C_j(0) = \frac{1}{\pi}\int^{\infty}_0d\omega\,J_j(\omega)\coth(\beta_j\omega/2)$, and 
\begin{equation}
	\fl \qquad \frac{\delta\tilde{\mathcal{F}}[\phi]}{\delta\phi_j(t)} = -\bigg(\int^t_0ds\,\tilde{\mathcal{B}}_j(t,s) + \int^t_0ds\,C_j(t-s)\phi_j(s)\bigg)\tilde{\mathcal{F}}[\phi] = -\Lambda_j[\phi](t)\tilde{\mathcal{F}}[\phi],
\end{equation}
resulting in
\begin{equation}\label{eq:}
    \frac{\delta^2\tilde{\rho}^B_S(\phi,t)}{\delta\phi_j(t)^2} = -\mathcal{T}\{[C_j(0) - \Lambda^2_j[\phi](t)]\tilde{\mathcal{F}}[\phi](t)\}\rho_S(0).
\end{equation}
By then inserting this expression into (\ref{eq:moments_fd}) with $n=2$, one obtains
\begin{equation}\label{eq:second_moment_gen}
	\langle I^2_{S_j}\rangle = -C_j(0){\rm Tr}\big[\tilde{S}^2_j(t)\tilde{\rho}_S(t)\big] + {\rm Tr}\big[\tilde{S}^2_j(t)\mathcal{T}\Lambda^2_j(t)\tilde{\mathcal{F}}(t)\rho_S(0)\big],
\end{equation}
with $\Lambda_j[\phi_j=0] = \Lambda_i(t)$. The quantity $\Lambda^2_j(t)$ may be expanded using the binomial theorem and substituted into (\ref{eq:second_moment_gen}) to give
\begin{equation}
	\fl \qquad \langle I^2_{S_j}\rangle = -C_j(0){\rm Tr}\big[S^2_j\rho_S(t)\big] - \sum_{k,k'}{\rm Tr}\big[S^2_j\big(\hat{\rho}_{\vec{0}^{+}_{jk,jk'},\vec{0}} + 2\hat{\rho}_{\vec{0}^+_{jk},\vec{0}^+_{jk'}} + \hat{\rho}_{\vec{0},\vec{0}^{+}_{jk,jk'}}\big)\big],
\end{equation}
where the $0_{jk,jk'}$ notation references second-tier ADOs with indices $0_{jk}$ and $0_{jk'}$ raised by one. 

Considering moments beyond second order, let us employ the following shorthand notation
\begin{equation}
	D^n_{\phi_j} \equiv \frac{\delta^n}{\delta\phi_j(t)^n}, \qquad D^0_{\phi_j}=1.
\end{equation}
The ($n+1$)-th derivative of $\tilde{\rho}^B_S(\phi,t)$ can be shown to satisfy
\begin{equation}\label{eq:MGF_rec}
	D^{n+1}_{\phi_j}\tilde{\rho}^B_S(\phi,t) = -\mathcal{T}\big\{nC_j(0)D^{n-1}_{\phi_j}\tilde{\mathcal{F}}[\phi] + \Lambda_j[\phi_j]D^n_{\phi_j}\tilde{\mathcal{F}}[\phi]\big\}\rho_S(0),
\end{equation}
where $D^n_{\phi_j}\tilde{\mathcal{F}}[\phi](t)$ obeys the recurrence relation
\begin{equation}\label{eq:FV_rec}
	D^{n+1}_{\phi_j}\tilde{\mathcal{F}}[\phi] = -nC_j(0)D^{n-1}_{\phi_j}\tilde{\mathcal{F}}[\phi] - \Lambda_j[\phi_j](t)D^n_{\phi_j}\tilde{\mathcal{F}}[\phi]. 
\end{equation}
Consequently, (\ref{eq:MGF_rec}) reduces to a sum of powers in $\Lambda_j[\phi_j]$ \cite{Song2017,Zhu2012},
\begin{equation}\label{eq:MGF_sum}
	D^{n+1}_{\phi_j}\tilde{\rho}^B_S(\phi,t) = \mathcal{T}\Bigg\{\Bigg(\sum^{n+1}_{\ell=0}b^{n+1}_{\ell}\Lambda^{\ell}_j[\phi_j]\Bigg)\tilde{\mathcal{F}}[\phi]\Bigg\}\rho_S(0),
\end{equation}
with
\begin{equation}
b^{n+1}_{\ell} = -nC_j(0)b^{n-1}_{\ell} -b^{n}_{\ell-1}, \qquad b^{0}_0=1.
\end{equation}
The terms $\Lambda^{\ell}_j[\phi_j]$ may then be expanded in a similar way to $\Lambda^2_j(t)$ to obtain $\langle I^n_{S_j}\rangle$ as a sum over ADOs. By doing so, one finds 
\begin{equation}
	\fl \qquad \langle I^{n+1}_{S_j}\rangle = \sum^{n+1}_{\ell=0}b^{n+1}_{\ell}i^{\ell}\sum^{\ell}_{\alpha=0}{\ell \choose \alpha}\sum_{k_1,...,k_{\ell-\alpha}}\sum_{k'_1,...,k'_{\alpha}}{\rm Tr}\Big[S^{n+1}_j\big(\hat{\rho}_{\vec{0}^+_{jk_1,...,_{jk_{m-\alpha}}},\vec{0}^+_{jk'_1,...,jk'_{\alpha}}}\big)\Big], 
\end{equation}
such that $n$-th order moments of the system current depend on ADOs up to the $n$-th level of the HEOM.

\subsection{Bath heat currents}

To calculate the bath heat current (\ref{eq:bhc_heom}) under the same approach, we insert $\tilde{\mathcal{O}}_j(t) = \tilde{A}_j(t)^+$ into (\ref{eq:moments}) with $A_j=i\sum_j\omega_kg_{k_j}(b^{\dagger}_{k_j} - b_{k_j})$. This allows us to write $I_{B_j}$ in terms of the generating functional $\tilde{\rho}^A_S(\phi,t)$,
\begin{equation}
	\langle I_{B_j}\rangle = i{\rm Tr}_S\left[\tilde{V}_j(t)\frac{\delta}{\delta\phi_j(t)}\tilde{\rho}^A_S(\phi,t)|_{\phi=0}\right].
\end{equation}
The evaluation of $\tilde{\rho}^A_S(\phi,t)$ follows the same procedure used above to derive a closed form expression for $\tilde{\rho}^B_S(\phi,t)$ \cite{Song2017}. After some algebra, we end up with
\begin{eqnarray}
	\fl \qquad \tilde{\rho}^A_S(\phi,t) &= \mathcal{T}\bigg\{\prod_j\exp\bigg[-\int^t_0ds\int^s_0du\Big(\phi_j(u)D_j(s-u)\tilde{V}_j(s)^- + \phi_j(s)\tilde{\mathcal{A}}_j(s,u)\nonumber \\
	&\qquad\qquad\quad\,\, + \phi_j(s)\phi_j(u)C'_j(s-u)\Big) - \int^t_0ds\,\tilde{\mathcal{W}}_j(s)\bigg]\bigg\}\rho_S(0), \\
	&\equiv \mathcal{T}\bigg\{\prod_j\exp\bigg[-\int^t_0ds\,\tilde{\mathcal{W}}'_j[\phi_j](s)\bigg]\bigg\}\rho_S(0), \nonumber
\end{eqnarray}
where 
\begin{equation}
	\tilde{\mathcal{A}}_j(t,s) = D_j(t-s)\tilde{V}_j(s)^+ - D^*_j(t-s)\tilde{V}_j(s)^-,
\end{equation}
and
\begin{eqnarray}
	\fl \quad D_j(t) &\equiv \frac{d}{dt}C_j(t) = \frac{i}{\pi}\int^{\infty}_0d\omega\,\omega J_j(\omega)\Big[i\coth\left(\frac{\beta_j\omega}{2}\right)\sin(\omega t) - \cos(\omega t)\Big], \label{eq:D}\\
	\fl \quad C'_j(t) &= \frac{1}{\pi}\int^{\infty}_0d\omega\,\omega^2J_j(\omega)\Big[\coth\left(\frac{\beta_j\omega}{2}\right)\cos(\omega t) -i\sin(\omega t)\Big]. 
\end{eqnarray}
The functional derivative of $\tilde{\rho}^A_S(\phi,t)$ may also be evaluated in a similar way to $\tilde{\rho}^B_S(\phi,t)$, resulting in
\begin{equation}\label{eq:func_derv_A}
	\fl \quad \frac{\delta}{\delta\phi_j(t)}\tilde{\rho}^A_S(\phi,t) = -\mathcal{T}\bigg\{\bigg[\int^t_0ds\,\tilde{\mathcal{A}}_j(t,s) + \int^t_0ds\,\phi_j(s)C'_j(t-s)\bigg]\tilde{\mathcal{F}}[\phi]\bigg\}\rho_S(0),
\end{equation}
such that we get the following closed expression for the bath heat current:
\begin{eqnarray}
	\fl \langle I_{B_j}\rangle &= -i{\rm Tr}_S\bigg[\tilde{V}_j(t)\int^t_0ds\mathcal{T}\{\tilde{\mathcal{A}}_j(t,s)\mathcal{F}(t)\}\rho_S(0)\bigg] \nonumber\\
	\fl	            &= -i\int^t_0ds\,D_j(t-s)\left\langle\mathcal{T}\,\tilde{V}_j(t)^+\tilde{V}_j(s)^+\tilde{\mathcal{F}}(t)\right\rangle_S + i\int^t_0ds\,D^*_j(t-s)\left\langle\mathcal{T}\,\tilde{V}_j(t)^+\tilde{V}_j(s)^-\tilde{\mathcal{F}}(t)\right\rangle_S. \nonumber
\end{eqnarray}
To now write $\langle I_{B_j}\rangle$ in terms of the ADOs, we can expand $D_j(t)$ and $D^*_j(t)$ into sums of exponentials using (\ref{eq:corr_func}) and (\ref{eq:D}). The overall result is
\begin{equation}\label{eq:bhc_ado}
	\langle I_{B_j}\rangle = -i\sum_kz_{jk}{\rm Tr}[V_j\hat{\rho}_{\vec{0}^+_{jk},\vec{0}}] + i\sum_kz^*_{jk}{\rm Tr}[V_j\hat{\rho}_{\vec{0},\vec{0}^+_{jk}}].
\end{equation}

Now, higher order moments of the bath heat current are determined from 
\begin{equation}
	\langle I^n_{B_j}\rangle = i^n{\rm Tr}_S\left[\tilde{V}^n_j(t)\frac{\delta^n}{\delta\phi_j(t)^n}\tilde{\rho}^A_S(\phi,t)|_{\phi=0}\right]. 
\end{equation}
which relies on evaluating functional derivatives $D^n_{\phi_j}\tilde{\rho}^A_S(\phi,t)$ for $n>1$. For this we proceed analogously to how functional derivatives $D^n_{\phi_j}\tilde{\rho}^B_S(\phi,t)$ were treated when evaluating the system currents. Specifically, we define
\begin{equation}
	\frac{\delta\tilde{\rho}^A_S(\phi,t)}{\delta\phi_j(t)} = -\mathcal{T}\{\Lambda'_j[\phi_j]\tilde{\mathcal{F}}'[\phi]\}\rho_S(0),
\end{equation}
where
\begin{equation}
	\Lambda'_j[\phi_j](t) = \int^t_0ds\,\big[\tilde{\mathcal{A}}_j(t,s) + \phi_j(s)C'_j(t-s)\big]
\end{equation}
and 
\begin{equation}
	\frac{\delta\Lambda'_j[\phi_j]}{\delta\phi_j(t)} = C'_j(0) = \frac{1}{\pi}\int^{\infty}_0d\omega\,\omega^2J_j(\omega)\coth(\beta_j\omega/2). 
\end{equation}
Hence, from (\ref{eq:func_derv_A}), 
\begin{equation}
	\frac{\delta\tilde{\mathcal{F}}'[\phi]}{\delta\phi_j(t)} = -\Lambda'_j[\phi_j](t)\tilde{\mathcal{F}}'[\phi], 
\end{equation}
one can show that $D^n_{\phi_j}\tilde{\rho}^A_S(\phi,t)$ satisfies the following recurrence relation
\begin{equation}\label{eq:MGF_rec}
	D^{n+1}_{\phi_j}\tilde{\rho}^A_S(\phi,t) = -\mathcal{T}\big\{nC'_j(0)D^{n-1}_{\phi_j}\tilde{\mathcal{F}}'[\phi] + \Lambda'_j[\phi_j]D^n_{\phi_j}\tilde{\mathcal{F}}'[\phi]\big\}\rho_S(0),
\end{equation}
and
\begin{equation}
	D^{n+1}_{\phi_j}\tilde{\mathcal{F}}'[\phi] = -nC'_j(0)D^{n-1}_{\phi_j}\tilde{\mathcal{F}}'[\phi] - \Lambda'_j[\phi_j]D^n_{\phi_j}\tilde{\mathcal{F}}'[\phi].
\end{equation}
Since these recurrence relations have an identical structure to (\ref{eq:MGF_sum}) and (\ref{eq:FV_rec}), then we may further write
\begin{equation}
	D^{n+1}_{\phi_j}\tilde{\rho}^A_S(\phi,t) = \mathcal{T}\Bigg\{\Bigg(\sum^{n+1}_{\ell=0}a^{n+1}_{\ell}\Lambda'^{\ell}_j[\phi_j]\Bigg)\tilde{\mathcal{F}}'[\phi]\Bigg\}\rho_S(0),
\end{equation}
where
\begin{equation}
	a^{n+1}_{\ell} = -nC'_j(0)a^{n-1}_{\ell} - a^n_{\ell-1}, \qquad a^0_0=1.
\end{equation}
Overall, the ($n+1$)-th order moments of the bath heat current thus read
\begin{eqnarray}
	\fl \qquad \langle I^{n+1}_{B_j}\rangle &= i^{n+1}\sum^{n+1}_{\ell=0}a^{n+1}_{\ell}\sum^{\ell}_{\alpha=0}{\ell \choose \alpha}(-1)^{\alpha}\sum_{k_1,...,k_{\ell-\alpha}}z_{jk_1}...z_{jk_{\ell-\alpha}}\sum_{k'_1,...,k'_{\alpha}}z^*_{jk'_1}...z^*_{jk'_{\ell-\alpha}} \nonumber\\
	\fl &\quad \times {\rm Tr}\Big[V^{n+1}_j\big(\hat{\rho}_{\vec{0}^+_{jk_1,...,jk_{m-\alpha}},\vec{0}^+_{jk'_1,...,jk'_{\alpha}}}\big)\Big],
\end{eqnarray}
where evidently the same relationship holds between the $n$-th order bath current $\langle I^n_{B_j}\rangle$ and ADOs as with $\langle I^n_{S_j}\rangle$.

\section{Fourth-order contribution to the heat current (\ref{eq:I_ss_pert})}\label{appen:C}

Starting from 
\begin{equation}\label{eq:shc_fourth_order}
I^{(2)}_{ss} = -i\lim_{t\rightarrow\infty}{\rm Tr}\Big\{H_S\big[\tilde{\sigma}_z(t),\tilde{\rho}^{(2)}_{\vec{0}^+_{1k},\vec{0}} + \tilde{\rho}^{(2)}_{\vec{0},\vec{0}^+_{1k}}\big]\Big\}, 
\end{equation}
we can eliminate $\tilde{\rho}^{(2)}_{\vec{0}^+_{1k},\vec{0}}$ ($\tilde{\rho}^{(2)}_{\vec{0},\vec{0}^+_{1k}}$) from this expression by solving the system of equations obtained from the HEOM evaluated at fourth order in $\lambda$,
\begin{eqnarray}\label{eq:ados_tier_2}
	\fl \qquad \qquad &\frac{d}{dt}\tilde{\rho}^{(2)}_{\vec{0}^+_{jk},\vec{0}} = -iz_{jk}\tilde{\rho}^{(2)}_{\vec{0}^+_{jk},\vec{0}} - i\sum_{\nu=1,2}\sum^{K_{\nu}}_{n=0}\tilde{V}_{\nu}(t)^{\times}\big[\tilde{\rho}^{(2)}_{\vec{0}^+_{\nu n,jk},\vec{0}} + \tilde{\rho}^{(2)}_{\vec{0}^+_{jk},\vec{0}^+_{\nu n}}\big], \\
	\fl \qquad \qquad &\frac{d}{dt}\tilde{\rho}^{(2)}_{\vec{0}^+_{jk},\vec{0}^+_{\nu n}} = -i(z_{jk} - z^*_{\nu n})\tilde{\rho}^{(2)}_{\vec{0}^+_{jk},\vec{0}^+_{\nu n}} - ic_{jk}\tilde{V}_j(t)^+\tilde{\rho}^{(1)}_{\vec{0},\vec{0}^+_{\nu n}}+ic^*_{\nu n}\tilde{V}_{\nu}(t)^-\tilde{\rho}^{(1)}_{\vec{0}^+_{jk},\vec{0}}, \nonumber \\
	\fl \qquad \qquad &\frac{d}{dt}\tilde{\rho}^{(2)}_{\vec{0}^+_{\nu n,jk},\vec{0}} = -i(z_{jk}+z_{\nu n})\tilde{\rho}^{(2)}_{\vec{0}^+_{\nu n,jk},\vec{0}}-ic_{jk}\tilde{V}_j(t)^+\tilde{\rho}^{(1)}_{\vec{0}^+_{\nu n},\vec{0}}-ic_{\nu n}\tilde{V}_{\nu}(t)^+\tilde{\rho}^{(1)}_{\vec{0}^+_{jk},\vec{0}}, \nonumber
\end{eqnarray}
with $O^{\times} \equiv O^{+} - O^{-}$. Note that here only first and second tier ADOs are retained, since those at higher levels (third tier and beyond) scale faster than $O(\lambda^4)$. Hence, we proceed by integrating the bottom two lines of (\ref{eq:ados_tier_2}) and inserting $\tilde{\rho}^{(1)}_{\vec{0}^+_{jk},\vec{0}}$ ($\tilde{\rho}^{(1)}_{\vec{0},\vec{0}^+_{jk}}$) from (\ref{eq:ados_tier_1_sol}) into the result to get
\begin{eqnarray}\label{eq:ados_tier_2_sol}
	\fl \qquad \tilde{\rho}^{(2)}_{\vec{0}^+_{jk},\vec{0}^+_{\nu n}}(t) &= \int^t_0ds\int^s_0du\Big[c_{jk}c^*_{\nu n}\tilde{V}_j(s)^+\tilde{V}_{\nu}(u)^-e^{-iz_{jk}(t-s)}e^{iz^*_{\nu n}(t-u)} \nonumber\\
	\fl \qquad &\qquad + c_{jk}c^*_{\nu n}\tilde{V}_{\nu}(s)^-\tilde{V}_j(u)^+e^{-iz_{jk}(t-u)}e^{iz^*_{\nu n}(t-s)}\Big]\tilde{\rho}_S(t) \nonumber\\ 
	\fl \qquad &\equiv -\int^t_0ds\,\tilde{\Upsilon}_{jk,\nu n}(t,s)\tilde{\rho}_S(t) \\
	\fl \qquad \tilde{\rho}^{(2)}_{\vec{0}^+_{\nu n,jk},\vec{0}}(t) &= -\int^t_0ds\int^s_0du\Big[c_{\nu n}c_{jk}\tilde{V}_{\nu}(s)^+\tilde{V}_j(u)^+e^{-iz_{jk}(s-u)}e^{-iz_{\nu n}(t-s)}  \nonumber\\ 
	\fl \qquad &\qquad + c_{\nu n}c_{jk}\tilde{V}_j(s)^+\tilde{V}_{\nu}(u)^+e^{-iz_{jk}(t-s)}e^{-iz_{\nu n}(t-u)}\Big]\tilde{\rho}_S(t) \nonumber\\
	\fl \qquad &\equiv -\int^t_0ds\,\tilde{\Phi}_{\nu n,jk}(t,s)\tilde{\rho}_S(t),
\end{eqnarray}
where above we have replaced $\tilde{\rho}_S(s)\rightarrow\tilde{\rho}_S(t)$, which is exact up to fourth-order given that 
\[
	\tilde{\rho}_S(s) = \tilde{\rho}_S(t) - \int^t_sdu\frac{d}{du}\tilde{\rho}_S(u) \equiv \tilde{\rho}_S(t) + O(\lambda^2).
\]
If we now formally integrate the top line of (\ref{eq:ados_tier_2}) and insert this into (\ref{eq:ados_tier_2_sol}), we obtain
\begin{equation}
      \fl \tilde{\rho}^{(2)}_{\vec{0}^+_{jk},\vec{0}}(t) = i\int^t_0ds\int^s_0du\sum_{\nu=1,2}\sum^{K_{\nu}}_{n=0}\tilde{V}^{\times}_j(s)\Big[\tilde{\Phi}_{\nu n,jk}(s,u) + \tilde{\Upsilon}_{jk,\nu n}(s,u)\Big]\tilde{\rho}_S(t)e^{-iz_{jk}(t-s)},
\end{equation}
which on substituting into (\ref{eq:shc_fourth_order}) gives
\begin{equation}
	I^{(2)}_{ss} = {\rm Tr}\{H_S\mathcal{D}^{(2)}_1[\rho^{ss}_S]\} 
\end{equation}
where 
\begin{eqnarray}
	\fl \mathcal{D}^{(2)}_j[\rho] &= \int^{\infty}_0dt_1\int^{\infty}_{t_1}dt_2\int^{\infty}_{t_2}dt_3\sum_{\nu=1,2}\Big(C_{\nu}(t_2-t_1)C_j(t_3)\big[V_j,[\tilde{V}_{\nu}(-t_1),\tilde{V}_{\nu}(-t_2)\tilde{V}_j(-t_3)\rho]\big] \nonumber\\ \fl & \qquad + C_j(t_2)C_{\nu}(t_3-t_1)\big[V_j,[\tilde{V}_{\nu}(-t_1),\tilde{V}_j(-t_2)\tilde{V}_{\nu}(-t_3)\rho]\big] \nonumber\\ \fl & \qquad - C_j(t_2)C^*_{\nu}(t_3-t_1)\big[V_j,[\tilde{V}_{\nu}(-t_1),\tilde{V}_j(-t_2)\rho\tilde{V}_{\nu}(-t_3)]\big] \nonumber\\ \fl & \qquad - C^*_{\nu}(t_2-1)C_j(t_3)\big[V_j,[\tilde{V}_{\nu}(-t_1),\tilde{V}_j(-t_3)\rho\tilde{V}_{\nu}(-t_2)]\big]\Big)+ {\rm h.c.} \label{eq:fod}
\end{eqnarray}
Notably, unlike the Redfield dissipator (\ref{eq:BRME_dis}), the above cannot be written in additive form---i.e. where $\mathcal{D}^{(2)}_j$ is made up of only local contributions associated with each bath $B_j$. Thus, $I^{(2)}_{ss}$ may account for non-additive heat flows in the nonequilibrium steady state.

The vectorized form of the dissipator (\ref{eq:fod}) reads
\begin{eqnarray*}
	& \fl {\rm vec}(\mathcal{D}^{(2)}_j[\rho]) = i(\mathbbm{1}\otimes V_j - V^T_j\otimes\mathbbm{1})\sum^{K_j}_{k=0}\sum_{\nu=1,2}\sum^{K_{\nu}}_{n=0}\sum_{\alpha_1,\alpha_2,\alpha_3}\sum_{\beta_1,\beta_2,\beta_3}(\mathbbm{1}\otimes V^{\alpha_1\beta_1}_{\nu}-V^{\alpha_1\beta_1\,T}_{\nu}\otimes\mathbbm{1})\\
	& \fl \qquad \times \bigg\{R^{jk,\nu n}_{\alpha_1\beta_1,\alpha_2\beta_2,\alpha_3\beta_3}(\mathbbm{1}\otimes V^{\alpha_2\beta_2}_{\nu}V^{\alpha_3\beta_3}_j) -R^{jk,(\nu n)^*}_{\alpha_1\beta_1,\alpha_2\beta_2,\alpha_3\beta_3}(V^{\alpha_2\beta_2\,T}_{\nu}\otimes V^{\alpha_3\beta_3}_j) \\
        & \fl \qquad+R^{(jk)^*,(\nu n)^*}_{\alpha_1\beta_1,\alpha_2\beta_2,\alpha_3\beta_3}(V^{\alpha_2\beta_2\,T}_{\nu}V^{\alpha_3\beta_3\,T}_j\otimes\mathbbm{1}) - R^{(jk)^*,\nu n}_{\alpha_1\beta_1,\alpha_2\beta_2,\alpha_3\beta_3}(V^{\alpha_3\beta_3\,T}_j\otimes V^{\alpha_2\beta_2}_{\nu})\bigg\}\rho
\end{eqnarray*}
where $V^{\alpha\beta}_{j} = \langle \alpha|V_j|\beta\rangle|\alpha\rangle\langle\beta|$ ($\alpha,\beta\in\{0,1\}$), 
\begin{eqnarray}
	& \fl R^{jk,\nu n}_{\alpha_1\beta_1,\alpha_2\beta_2,\alpha_3\beta_3} = \frac{c_{jk}c_{\nu n}}{(\omega_{\alpha_3\beta_3}+z_{jk})(\omega_{\alpha_2\beta_2}+\omega_{\alpha_3\beta_3}+z_{jk}+z_{\nu n})(\omega_{\alpha_1\beta_1}+\omega_{\alpha_2\beta_2}+\omega_{\alpha_3\beta_3}+z_{jk})} \nonumber\\
	& \fl \qquad + \frac{c_{jk}c_{\nu n}}{(\omega_{\alpha_3\beta_3}+z_{\nu n})(\omega_{\alpha_2\beta_2}+\omega_{\alpha_3\beta_3}+z_{\nu n}+z_{jk})(\omega_{\alpha_1\beta_1}+\omega_{\alpha_2\beta_2}+\omega_{\alpha_3\beta_3}+z_{jk})} \label{eq:R}
\end{eqnarray}
and the notation $(jk)^*$ indicates the replacement $c_{jk},z_{jk}\rightarrow c^*_{jk},-z^*_{jk}$ in expression (\ref{eq:R}).

\end{document}